\documentstyle{article}
\include{psfig}

\title{Neural Network Design for J Function Approximation in Dynamic Programming}

\author{Xiaozhong Pang \and Paul J. Werbos}
\date{}
\begin{document}
\maketitle
\thispagestyle{empty}
\begin{abstract}
This paper will show that a new neural network design can solve an 
example of difficult function approximation problems which are crucial 
to the field of approximate  dynamic programming(ADP). Although 
conventional neural networks have been proven to approximate smooth functions very 
well, the use of ADP for problems of intelligent control or planning 
requires the approximation of functions which are not so smooth. As an
example, this paper studies the problem of approximating the $J$ function 
of dynamic programming applied to the task of navigating mazes in general 
without the need to learn each individual maze. Conventional neural networks, 
like multi-layer perceptrons(MLPs), cannot learn this task. But a new type 
of neural networks, simultaneous recurrent networks(SRNs), can do so
as demonstrated by successful initial tests. The paper also examines the ability of recurrent 
neural networks to approximate MLPs and vice versa.

\end{abstract}

{\bf Keywords:} Simultaneous recurrent networks(SRNs), multi-layer
 perceptrons(MLPs), approximate dynamic programming, maze navigation, neural networks.

\section{Introduction}

\subsection{Purpose}

This paper has three goals: 

First, to demonstrate the value of a new class of neural network which provides a 
crucial component needed for brain-like intelligent control systems for the future.

Second, to demonstrate that this new kind of neural network provides better function 
approximate ability for use in more ordinary kinds of neural network applications for 
supervised learning.

Third, to demonstrate some practical implementation techniques necessary to make this 
kind of network actually work in practice.

\begin{figure}[htbp] 
\centerline{\psfig{figure=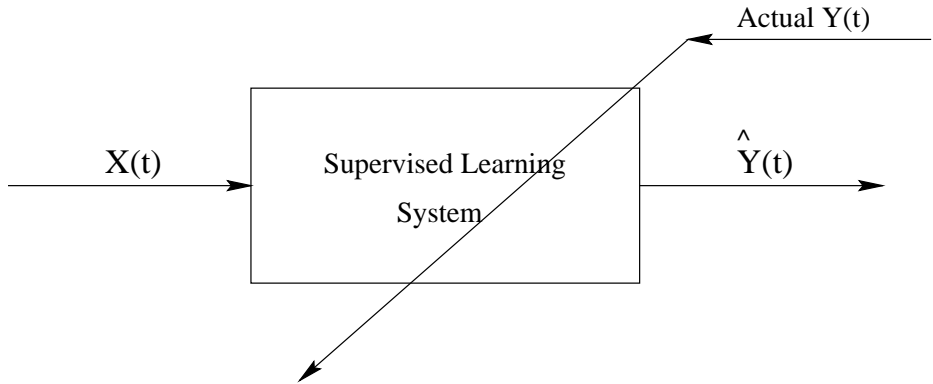,height=2.0in}}
\caption{What is supervised learning?}
\label{channel}
\end{figure}
 
\subsection{Background}
At present, in the neural network field perhaps $90\%$ of neural
network applications involve the use of neural networks designed to
performance a task called supervised learning(Figure 1). Supervised
learning is the task of learning a nonlinear function which may have
several inputs and several outputs based on some examples of the
function. For example, in character recognition, the inputs may be an
array of pixels seen from a camera. The desired outputs of the network
may be a classification of character being seen. Another
example would be for intelligent sensing in the chemical industry where
the inputs might be spectral data from observing a batch of chemicals,
and the desired outputs would be the concentrations of the different
chemicals in the batch. The purpose of this application is to predict or estimate what is in the batch without the need for expensive analytical tests.

The work in this paper will focus totally on certain tasks in
supervised learning. Even though existing neural networks can be used
in supervised learning, there can be performance problems depending on what kind of function
is learned. Many people have proven many theorems to show that
neural networks, fuzzy logic, Taylor theories and other function
approximation have a universal ability to approximate functions on the
condition that the functions have certain properties and that there is no limit on the complexity of the approximation. In practice, many approximation schemes become useless when there are many input variables because the required complexity grows at an exponential rate. 

For example, one way to approximate a function would be to construct a table of the values of the function at certain points in the space of possible inputs. Suppose there are 30 input variables and we consider 10 possible values of each input. In that case, the table must have $10^{30}$ numbers in it. This is not useful in practice for many reasons. Actually, however, many popular approximation methods like radial basis function(RBF) are similar in spirit to a table of values.

In the field of supervised learning, Andrew Barron[30] has proved some
function approximation theorems which are much more useful in
practice. He has proven that the most popular form of neural networks,
the multi-layer perceptron(MLP), can approximate any smooth
function. Unlike the case with the linear basis functions (like RBF
and Taylor series), the complexity of the network does not grow
rapidly as the number of input variables grows.

Unfortunately there are many practical applications where the functions
to be approximated are not smooth. In some cases, it is good enough
just to add extra layers to an MLP[1] or to use a generalized
MLP[2]. However, there are some difficult problems which arise in
fields like intelligent control or image processing or even stochastic
search where feed-forward networks do not appear powerful enough. 

\subsection{Summary and Organization of This Paper}
The main goal of this paper is to demonstrate the capability of a
different kind of supervised learning system based on a kind of
recurrent network called simultaneous recurrent network(SRN). In the
next chapter we will explain why this kind of improved supervised
learning system will be very important to intelligent control and to
approximate dynamic programming. In effect this work on supervised
learning is the first step in a multi-step effort to build more
brain-like intelligent systems. The next step would be to apply the SRN to
static optimization problems, and then to integrate the SRNs into large
systems for ADP. 

Even though intelligent control is the main motivation for this work,
the work may be useful for other areas as well. For example, in zip
code recognition, $AT\&T$[3] has demonstrated that feed-forward networks
can achieve a high level of accuracy in classifying individual
digits. However, $AT\&T$ and the others still have difficulty in
segmenting the total zip codes into individual digits. Research on
human vision by von der Malsburg[4] and others has suggested that some
kinds of recurrency in neural networks are crucial to their abilities in
image segmentation and binocular vision. Furthermore, researchers in
image processing like Laveen Kanal have showed that iterative
relaxation algorithms are necessary even to achieve moderate success in such
image processing tasks. Conceptually the SRN can learn an optimal
iterative algorithm, but the MLP cannot represent any iterative algorithms. In summary, though we are most interested in brain-like intelligent control, the development of SRNs could lead to very important applications in areas such as image processing in the future.

The network described in this paper is unique in several
respects. However, it is certainly not the first serious use of a
recurrent neural network. Chapter 3 of this paper will describe the
existing literature on recurrent networks. It will describe the
relationship between this new design and other designs in the
literature. Roughly speaking, the vast bulk of research in recurrent
networks has been academic research using designs based on ordinary
differential equations(ODE) to perform some tasks very different from
supervised learning --- tasks like clustering, associative memory and
feature extraction. The simple Hebbian learning methods[13] used for those
tasks do not lead to the best performance in supervised learning. Many
engineers have used another type of recurrent network , the time
lagged recurrent network(TLRN), where the recurrency is used to
provide memory of past time periods for use in forecasting the
future. However, that kind of recurrency cannot provide the iterative
analysis capability mentioned above. Very few researchers have written
about SRNs, a type of recurrent network designed to minimize error and
learn an optimal iterative approximation to a function. This is
certainly the first use of SRNs to learn a $J$ function from dynamic programming which will be explained more in chapter 2. This may also be the first empirical demonstration of the need for advanced training methods to permit SRNs to learn difficult functions.

Chapter 4 will explain in more detail the two test problems we have
used for the SRN and the MLP, as well as the details of architecture and learning procedure.

The first test problem was used mainly as an initial test of a simple
form of SRNs. In this problem, we tried to test the hypothesis that an
SRN can always learn to approximate a randomly chosen MLP, but not
vice versa. Although our results are consistent with that hypothesis,
there is room for more extensive work in the future, such as experiments with different sizes of neural networks and more complex statistical analysis.

The main test problem in this work was the problem of learning the $J$
function of dynamic programming. For a maze navigation problem, many
neural network researchers have written about neural networks which
learn an optimal policy of action for one particular maze[5]. This paper will address the more difficult problem of training
a neural network to input a picture of a maze and output the $J$
function for this maze. When the $J$ function is known, it is a
trivial local calculation to find the best direction of movement. This
kind of neural network should not require retraining whenever a new
maze is encountered. Instead it should be able to look at the maze and
immediately "see" the optimal strategy. Training such a network is a
very difficult problem which has never been solved in the past with
any kind of neural network. Also it is typical of the challenges one
encounters in true intelligent control and planning. This paper has demonstrated a working solution to this problem for the first time. Now that a system is working on a very simple form for this problem, it would be possible in the future to perform many tests of the ability of this system to generalize its success to many mazes.

In order to solve the maze problem, it was not sufficient only to use
an SRN. There are many choices to make when implementing the general
idea of SRNs or MLPs. Chapter 5 will describe in detail how these choices were 
made in this work. The most important choices were:

1. Both for the MLP and for the feed-forward core of the SRN we used the generalized MLP design[2] which eliminates the need to decide on the number of layers.

2. For the maze problem, we used a cellular or weight-sharing
architecture which exploits the spatial symmetry of the problem and
reduces dramatically the number of weights. In effect we solved the
maze problem using only five distinct neurons. There are interesting
parallels between this network and the hippocampus of the human brain.

3. For the maze problem, an adaptive learning rate(ALR) procedure was used to prevent oscillation and ensure convergence.

4. Initial values for the weights and the initial input vector for the SRN were chosen essentially at random, by hand. In the future, more systematic methods are available. But this was sufficient for success in this case.

Finally Chapter 6 will discuss the simulation results in more detail, give the conclusions of this paper and mention some possibilities for future work.

\section{Motivation}

In this chapter we will explain the importance of this work. As discussed above, the paper shows how to use a new type of neural
network in order to achieve better function approximation than what
is available from the types of neural networks which are popular
today. This chapter will try to explain why better function
approximation is important to approximate dynamic programming(ADP),
intelligent control and understanding the brain. Image processing and
other applications have already been discussed in the Introduction. These
three topics --- ADP, intelligent control and understanding
the brain --- are all closely related to each other and provide the original
motivation for the work of this paper.

The purpose of this paper is to make a core contribution to developing
the most powerful possible system for intelligent control.

In order to build the best intelligent control systems, we need to
combine the most suitable mathematics together with some understanding
of natural intelligence in the brain. There is a lot of interest
in intelligent control in the world. Some control systems which are
called intelligent are actually very quick and easy
things. There are many people who try to move step by step to add intelligence
into control , but a step-by-step approach may not be enough by itself.

Sometimes to achieve a complex difficult goal, it is necessary to have
a plan, thus some parts of the intelligent control community have
developed a
more systematic vision or plan for how it could be possible to achieve
real intelligent control. First, one must think about the question of
what is intelligent control. Then, instead of trying to answer this
question in one step, we try to develop a plan to reach the
design. Actually there are two questions:

1. How could we build an artificial system which replicates the main
capabilities of brain-like intelligence, somehow unified together as they are
unified together in the brain?

2. How can we understand what are the capabilities in the brain and how
they are organized in a functional engineering view? i.e. how are those
circuits in the human brain arranged to learn how to perform different
tasks? 

It would be best to understand how the human brain works before
building an artificial system. However, at the present time, our
understanding of the brain is limited. But at least we know that local
recurrency plays critical rule in the higher part of the human
brain[6][7][8][4].

Another reason to use SRNs is that SRNs can be very useful in ADP
mathematically. Now we will discuss what ADP can do for intelligent
control and understanding the brain. 

The remainder of this chapter will address three questions in order:\\
1. What is ADP?\\
2. What is the importance of ADP to intelligent control and
understanding the brain?\\
3. What is the importance of SRNs to ADP?

\subsection{What is ADP and J Function?}

To explain what is ADP, let us consider the original Bellman equation[9]:

\begin{equation}
J(R(t))={\max_{u(t)}(U(R(t),u(t))+<J(R(t+1))>)}/(1+r)-U_{0}
\end{equation}                                      
where $r$ and $u_{0}$ are constants that are used only in
infinite-time-horizon problems and then only sometimes, and where the
angle brackets refer to expectation value. In this paper we actually use:

\begin{equation}
J(R(t))=\max_{u(t)}(U(R(t),u(t))+<J(R(t+1))>)
\end{equation}
since the maze problem do not involve an infinite time-horizon.

Instead of solving for the value of $J$ in every possible state,
$R(t)$, we can use a function approximation method like neural networks
to approximate the $J$ function. This is called approximate dynamic programming(ADP). This paper
is not doing true ADP because in true ADP we do not know what the $J$
function is and must therefore use indirect methods to approximate
it. However, before we try to use SRNs as a component of an ADP
system, it makes sense to first test the ability of an SRN to
approximate a $J$ function, in principle.

Now we will try to explain what is the intuitive meaning of the Bellman
equation(Equation(1)) and the $J$ function according to the treatment taken
from[2].

To understand ADP, one must first review the basics of classical
dynamic programming, especially the versions developed by Howard[28] and
Bertsekas. Classical dynamic programming is the only exact and
efficient method to compute the optimal control policy over time, in a
general nonlinear stochastic environment. The only reason to approximate it
is to reduce computational cost, so as to make the method affordable
(feasible) across a wide range of applications. In dynamic programming, the user supplies a utility function which
may take the form $U(R(t),u(t))$ --- where the vector R is a Representation or
estimate of the state of the environment (i.e. the state vector) --- and a
stochastic model of the plant or environment. Then "dynamic programming"
(i.e. solution of the Bellman equation) gives us back a secondary or
strategic utility function $J(R)$. The basic theorem is that maximizing
$U(R(t),u(t))+J(R(t+1))$ yields the optimal strategy, the policy which will
maximize the expected value of $U$ added up over all future time. Thus
dynamic programming converts a difficult problem in optimizing over many
time intervals into a straightforward problem in short-term maximization.
In classical dynamic programming, we find the exact function $J$ which
exactly solves the Bellman equation. In ADP, we learn a kind of "model" of
the function $J$; this "model" is called a "Critic." (Alternatively, some
methods learn a model of the derivatives of $J$ with respect to the variables
$R_{i}$ ; these correspond to Lagrange multipliers, $\lambda_{i}$ , and to the "price variables" of microeconomic theory. Some methods learn a function related
to $J$, as in the Action-Dependent Adaptive Critic (ADAC)[29].

\subsection{Intelligent Control and Robust Control}

To understand the human brain scientifically, we must have some
suitable mathematical
concepts. Since the human brain makes decisions like a control system, it
is an example of an intelligent control system. Neuroscientists do not
yet understand the general ability of the human brain to learn to
perform new tasks and solve new problems even though they have studied
the brain for decades. Some people compare the past research in this
field to what would happen if we spent years to study
radios without knowing the mathematics of signal processing. 

We first
need some mathematical ideas of how it is possible for a computing
system to have this kind of capability based on distributed parallel
computation. Then we must ask what are the most important abilities of
the human brain which unify all of its more specific abilities in
specific tasks. It would be seen that the most important ability of brain
is the ability to learn over time how to make better decisions in
order to better maximize the goals of the organism. The natural way to
imitate this capability in engineering systems is to build systems which
learn over time how to make decisions which maximize some measure of
success or utility over future time. In this context, dynamic
programming is important because it is the only exact and efficient 
method for maximizing utility over future time. In the general situation,
where random disturbances and nonlinearity are expected, ADP is
important because it provides both the learning capability and the
possibility of reducing computational cost to an affordable level. For
this reason, ADP is the only approach we have to imitating this kind of
ability of the brain.

The similarity between some ADP designs and the circuitry of the
brain has been discussed at length in [10] and [11]. For example, there is
an important structure in the brain called the limbic system which
performs some kinds of evaluation or reinforcement functions, very
similar to the functions of the neural networks that must approximate
the $J$ function of dynamic programming. The largest part of the limbic
system, called the hippocampus, is known to possess a higher degree of
local recurrency[8].

In general, there are two ways to make classical controllers stable
despite great uncertainty about parameters of the plant to be
controlled. For example, in controlling a high speed aircraft, the
location of the center of the gravity is not known. The center of gravity is not known
exactly because it depends on the cargo of the air plane and the
location of the passengers. One way to account for such uncertainties is to
use adaptive control methods. We can get similar results, but more
assurance of stability in most cases[16] by using related neural
network methods, such as adaptive critics with recurrent networks. It is like
adaptive control but more general. There is another approach called robust
control or $H_{\infty}$ control, which trys to design a fixed controller
which remains stable over a large range in parameter space. Baras and
Patel[31] have for the first time solved the general problem of
 $H_{\infty}$ control for general partially observed nonlinear plants. They
have shown that this problem reduces to a problem in nonlinear, stochastic
optimization.  Adaptive dynamic programming makes it possible to solve
large scale problems of this type.

\subsection{Importance of the SRN to ADP}

ADP systems already exist which perform relatively simple control
tasks like stabilizing an aircraft as it lands under windy conditions
[12]. However this kind of task does not really
represent the highest level of intelligence or planning. True
intelligent control requires the ability to make decisions when future
time periods will follow a complicated, unknown path starting from the initial state. One example of a
challenge for intelligent control is the problem of navigating a
maze which we will discuss in chapter 4. A true intelligent control system
should be able to learn this kind of task. However, the ADP systems in
use today could never
learn this kind of task. They use conventional neural networks to approximate the
$J$ function. Because the conventional MLP cannot
approximate such a $J$ function, we may deduce that ADP system
constructed only from MLPs will never be able to display this kind of
intelligent control. Therefore, it is essential that we can find a kind
of neural network which can perform this kind of task. As we will
show, the SRN can fill this crucial gap. There are additional reasons for
believing that the SRN may be crucial to intelligent control as discussed
in chapter 13 of [9].

\section{Alternative Forms of Recurrent Networks} 

\subsection{Recurrent Networks in General}

There is a huge literature on recurrent networks. Biologists have used
many recurrent models because the existence of recurrency in the brain is
obvious. However, most of the recurrent networks implemented so far
have been classic style recurrent networks, as shown on the left hand
of Figure 2. Most of these networks are formulated from ordinary
differential equation(ODE) systems. Usually their learning is based on a
restricted concept of Hebbian learning. Originally in the neural
network field, the most popular neural networks were recurrent
networks like those which Hopfield[14] and Grossberg[15] used to provide
associative memory. 

\begin{figure}[htbp] 
\centerline{\psfig{figure=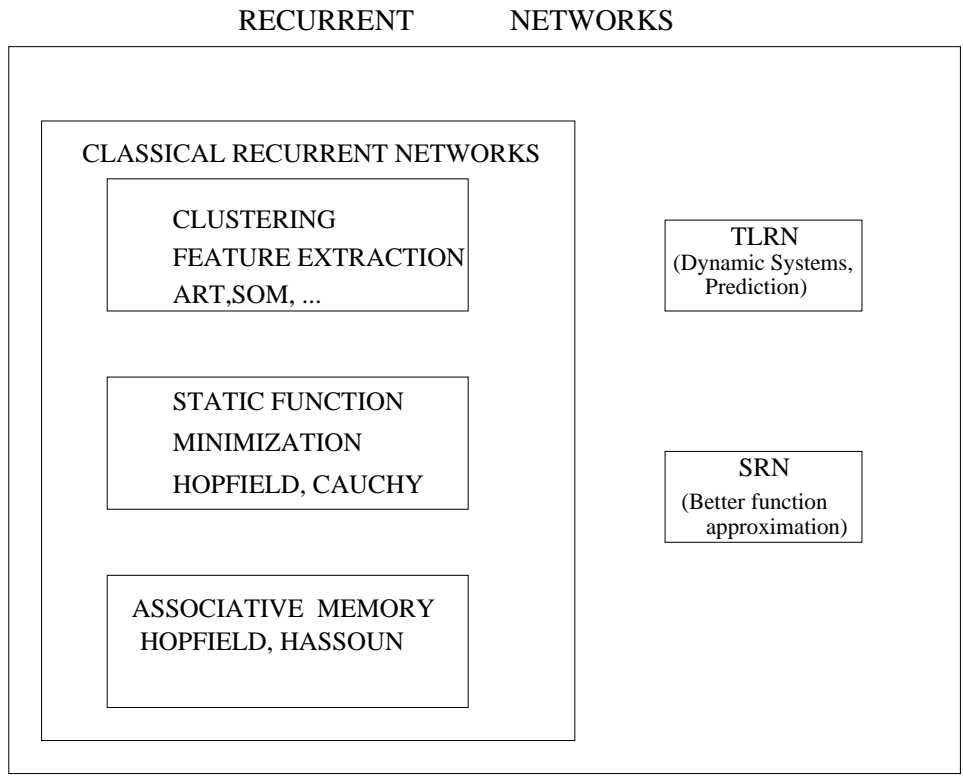,height=4.0in}}
\caption{Recurrent networks}
\label{channel}
\end{figure}

Associative memory networks can actually be applied
to supervised learning. But in actuality their capabilities are very
similar to those of look-up tables and radial basis functions. They make
predictions based on similarity to previous examples or
prototypes. They do not really try to estimate general functional
relationships. As a result these methods have become unpopular in
practical applications of supervised learning. The theorems of Barron
discussed in the Introduction show that MLPs do provide better
function approximation than do simple methods based on similarity.

There has been substantial progress in the past few years in
developing new associative memory designs. Nevertheless, the MLP is still
better for the specific task of
function approximation which is the focus of this paper.

In a similar way, classic recurrent networks have been used for tasks
like clustering, feature extraction and static function
optimization. But these are different problems from what we are trying
to solve here.

Actually the problem of static optimization will be considered in
future stages of this research. We hope that the SRN can be useful in such
applications after we have used it for supervised learning. When people use
the classic Hopfield networks for static optimization, they specify
all the weights and connections in advance[14]. This has limited the
success of this kind of network for large scale problems where it is difficult
to guess the weights. With the SRN we have methods to train the
weights in that kind of structure. Thus the guessing is no longer
needed. However, to use SRNs in that application requires refinement
beyond the scope of this paper. 

There have also been researchers using
ODE neural networks who have tried to use training schemes based on a
minimization of error instead of Hebbian approaches. However, in
practical applications of such networks, it is important to consider
the clock rates of computation and data sampling. For that reason, it
is both easier and better to use error minimizing designs based on
discrete time rather than ODE.

\subsection{Structure of Discrete-Time Recurrent Networks}

If the importance of neural networks is measured by the number of
words published, then the classic networks dominate the field of
recurrent networks. However, if the value is measured based on
economic value of practical application, then the field is dominated
by time-lagged recurrent networks(TLRNs). The purpose of the TLRN
is to predict or classify time-varying systems using recurrency as a
way to provide memory of the past. The SRN has some relation with the TLRN but
it is designed to perform a fundamentally different task. The SRN
uses recurrency to represent more complex relationships between one
input vector $X(t)$ and one output $Y(t)$ without consideration of the
other times $t$. Figure 3 and Figure 4 show us more details about
the TLRN and the SRN.

\begin{figure}[htbp] 
\centerline{\psfig{figure=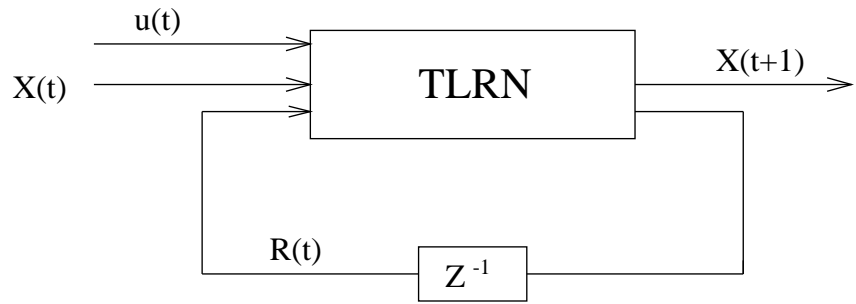,height=2.0in}}
\caption{Time lagged recurrent network(TLRN)}
\label{channel}
\end{figure}

\begin{figure}[htbp] 
\centerline{\psfig{figure=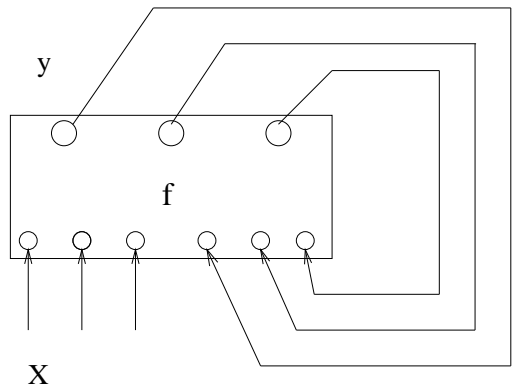,height=2.0in}}
\caption{Simultaneous recurrent network(SRN)}
\label{channel}
\end{figure}

In control applications, $u(t)$ represents the control variables which
we use to control the plant. For example, if we design a controller for
a car engine, the $X(t)$ variables are the data we get from our
sensors. The $u(t)$ variables would include the valve settings which we
use to try to control the process of combustion. The $R(t)$ variables
provide a way for the neural networks to remember past time cycles,
and to implicitly estimate important variables which cannot be observed
directly. In fact, the application of TLRNs to automobile control is
the most valuable application of recurrent networks ever developed so far.

A simultaneous recurrent network(Figure 4) is defined as a mapping:

\begin{equation}
{{\hat{Y}}(t)}=F(X(t),W)
\end{equation}

which is computed by iterating over the following equation:

\begin{equation}
{{y^{(n+1)}}(t)}= f({y^{(n)}}(t),X(t),W)
\end{equation}
where $f$ is some sort of feed-forward network or system, and $\hat{Y}$ is defined
as:

\begin{equation}
{{\hat{Y}}(t)}={\lim_{n \rightarrow \infty} y^{(n)}(t)}
\end{equation}

When we use $\hat{Y}$ in this paper, we use $n=20$ instead of $\infty$ here.

In Figure 4, the outputs of the neural network come back again as inputs
to the same network. However, in concept there is no time
delay. The inputs and outputs should be simultaneous. That is why it
is called a simultaneous recurrent network(SRN). In practice, of course,
there will always be some physical time delay between the outputs and
the inputs. However if the SRN is implemented in fast computers, this
time delay may be very small compared to the delay between different
frames of input data. 

In Figure 4, $X$ refers to the input data at the current time frame $t$. The
vector $y$ represents the temporary output of the network, which is then
recycled as an additional set of inputs to the network. At the center
of the SRN is actually the feed-forward network which implements the
function $f$. (In designing an SRN, you can choose any feed-forward
network or system as you like. The function $f$ simply describes which
network you use). The output of the SRN at any time $t$ is simply the limit
of the temporary output $y$.

In Equation (3) and (4), notice that there are two integers --- $n$ and $t$ ---
which could both represent some kind of time. The integer $t$
represents a slower kind of time cycle, like the delay between frames
of incoming data. The integer $n$ represents a faster kind of time,
like the computing cycle of a fast electronic chip. For example, if we
build a computer to analyze images coming from a movie camera, "$t$" and
"$t+1$" represent two successive incoming pictures with a movie
camera. There are usually only 32 frames per second. (In the human brain,
it seems that  there are only about 10 frames per second coming into
the neocortex.) But
if we use a fast neural network chip, the computational cycle ---
the time between "$n$" and "$n+1$" --- could be as small as a microsecond.

In actuality, it is not necessary to choose between time-lagged
recurrency (from $t$ to $t+1$) and simultaneous recurrency (from $n$ to
$n+1$). It is possible to build a hybrid system which contains both
types of recurrency. This could be very useful in analyzing data like
movie pictures, where we need both memory and some ability to segment
the images. [9] discusses how to build such a hybrid. However,
before building such a hybrid, we must first learn to make SRNs work
by themselves.

Finally, please note that the TLRN is not the only kind of neural
network used in predicting dynamical systems. Even more popular is
the time delayed neural network(TDNN), shown in Figure 5. The TDNN is popular because it is easy to use.
However, it has less capability, in principle, because it has no
ability to estimate unknown variables. It is especially weak when some
of these variables change slowly over time and require memory which
persists over long time periods. In addition, the TLRN fits the
requirements of ADP directly, while the TDNN does not[9][16].

\begin{figure}[htbp] 
\centerline{\psfig{figure=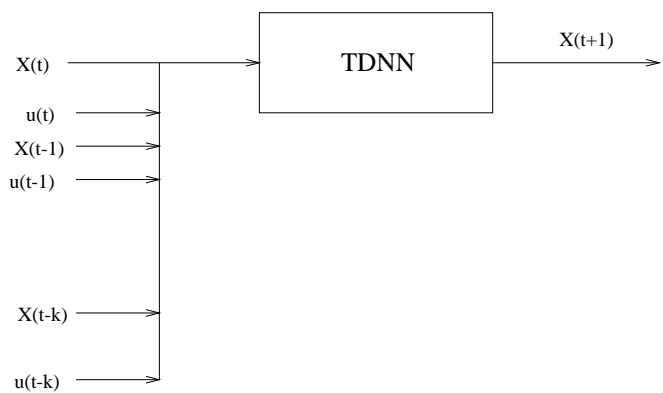,height=2.0in}}
\caption{Time delayed neural network(TDNN)}
\label{channel}
\end{figure}

\subsection{Training of SRNs and TLRNs}

There are many types of training that have been used for recurrent
networks. Different types of training give rise to different kinds of
capabilities for different tasks. For the tasks which we have described for
the SRN and the TLRN, the proper forms of training all involve some calculation
of the derivatives of error with respects to the weights. Usually
after these derivatives are known, the weights are adapted according
to a simple formula as follows:

\begin{equation}
new W_{i,j} = old W_{i,j} - LR*{\frac{\partial Error}{\partial W_{i,j}}}
\end{equation}

where $LR$ is called the learning rate.

There are five main ways to train SRNs, all based on different methods
for calculating or approximating the derivatives. Four of these
methods can also be used with TLRNs. Some can be used for control
applications. But the details of those applications are beyond the
scope of this paper. These five types of training are listed in
Figure 6. For this paper, we have used two of these methods: Backpropagation
through time(BTT) and Truncation. 

\begin{figure}[htbp] 
\centerline{\psfig{figure=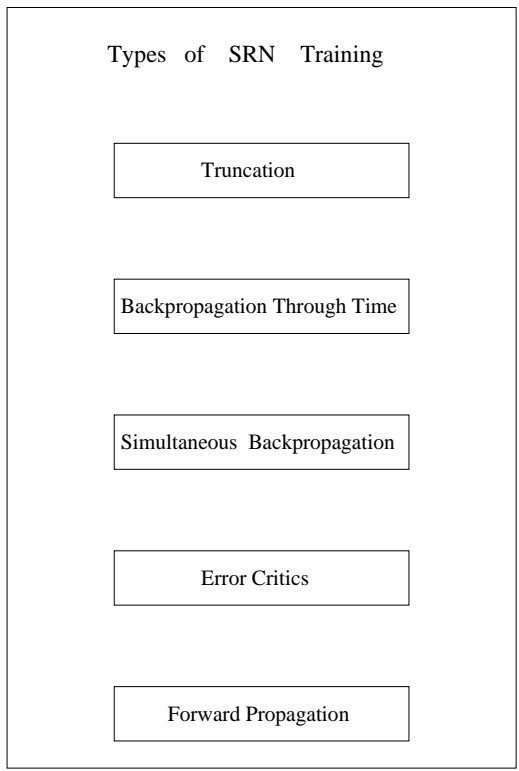,height=4.0in}}
\caption{Types of SRN Training}
\label{channel}
\end{figure}

The five methods are:

1. Backpropagation through time(BTT). This method and forward
propagation are the two methods which calculate the derivatives
exactly. BTT is also less expensive than forward propagation.

2. Truncation. This is the simplest and least expensive method. It uses
only one simple pass of backpropagation through the last iteration of
the model. Truncation is probably the most popular method used to adapt SRNs
even though the people who use it mostly just call it ordinary
backpropagation.

3. Simultaneous backpropagation. This is more complex than
truncation, but it still can be used in real time learning. It calculates
derivatives which are exact in the neighborhood of equilibrium but it does
not account for the details of the network before it reaches the
neighborhood of equilibrium.

4. Error critics(shown in Figure 7). This provides a general approximation to BTT which is
suitable for use in real-time learning[9].

\begin{figure}[htbp] 
\centerline{\psfig{figure=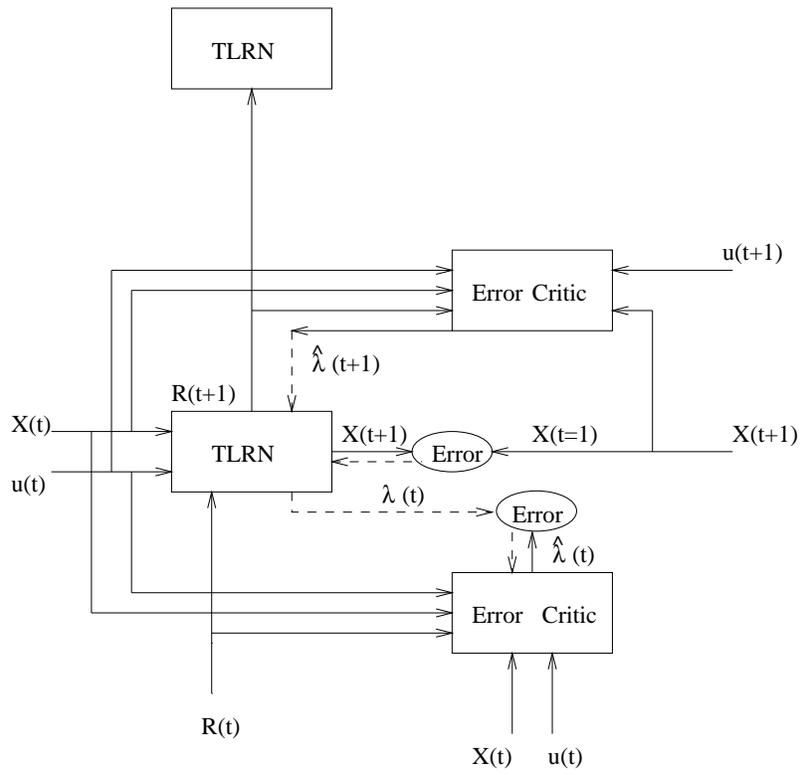,height=4.0in}}
\caption{Error Critics}
\label{channel}
\end{figure}

5. Forward propagation. This, like BTT, calculates exact derivatives. It is
often considered suitable for real-time learning because the
calculations go forward in time. However, when there are $n$ neurons and
$m$ connections, the cost of this method per unit of time is
proportional to $n*m$. Because of this high cost, forward propagation is
not really brain-like any more than BTT. 

\subsubsection{Backpropagation through time(BTT)}

BTT is a general method for calculating all the derivative of any
outcome or result of a process which involves repeated calls to the same
network or networks used to help calculate some kind of final outcome
variable or result $E$. In some applications, $E$ could represent
utility, performance, cost or other such variables. But in this
paper, $E$ will be used to represent error. BTT was first proposed
and implemented in [17]. The general form of BTT is as follows:
\\
for ${k = 1}$ to $T$ do forward$\_$calculation($k$);\\
     calculate result $E$;\\
     calculate direct derivatives of $E$ with respect to outputs of forward
               calculations;\\
     for ${k = T}$ to 1 backpropagate through
     forwards$\_$calculation($k$), calculating running totals where appropriate.

\begin{figure}[htbp] 
\centerline{\psfig{figure=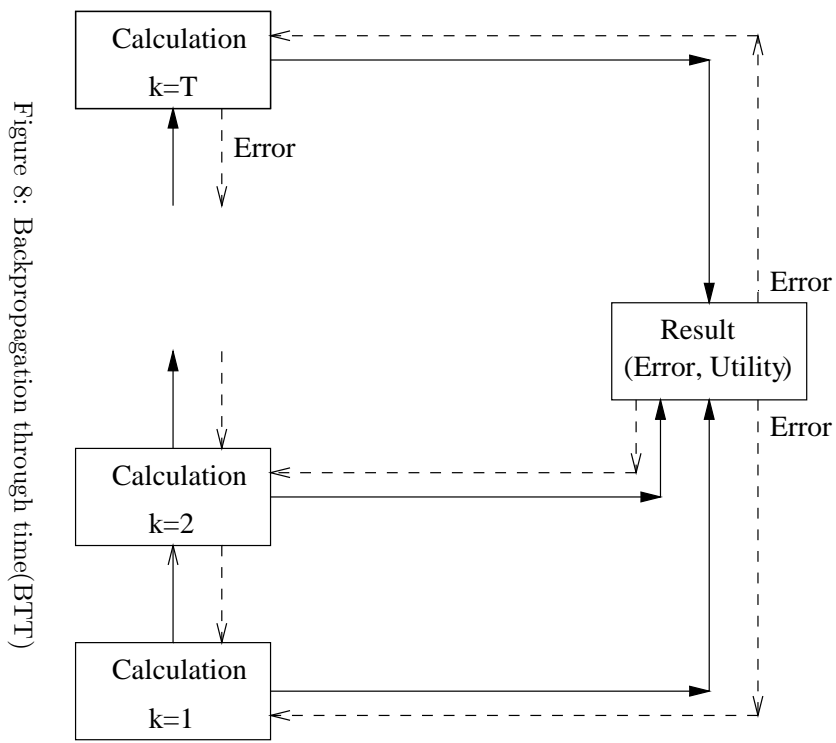,height=4.0in}}
\caption{Backpropagation through time(BTT)}
\label{channel}
\end{figure}

These steps are illustrated in Figure 8. Notice that this algorithm can
be applied to all kinds of calculations. Thus we can apply it
to cases where $k$ represents data frames $t$ as in the TLRNs, or to cases where $k$
represents internal iterations $n$ as in the SRNs. Also note that each box of
calculation receives input from some dashed lines which represent the
derivatives of $E$ with respect to the output of the box. In
order to calculate the derivatives coming out of each calculation box,
one simply uses backpropagation through the calculation of that box
starting out from the incoming derivatives. We will explain in more
detail how this works in the SRN case and the TLRN case.

So far as we know BTT has been applied in published working systems for
TLRNs and for control, but not yet for SRNs until now. However,
Rumelhart, Hinton and Williams[18] did suggest that someone should try
this.

The application of BTT for TLRNs is described at length in [2] and
[9]. The procedure is illustrated in Figure 9. In this example the
total error is actually the sum of error over each time $t$ where $t$ goes
from 1 to $T$. Therefore the outputs of the TLRN at each time $t$ $(t<T)$ have two ways of changing total errors:

(1)A direct way when the current predictions $\hat{Y}(t)$ are different from
the current targets $Y(t)$;

(2)An indirect way based on the impact of $R(t)$ on errors in later time
periods.

Therefore the derivative feedback coming into the TLRN is actually the
sum of two feedbacks from two different sources. As a technical detail,
note that $R(0)$ needs to be specified somehow. However, we will not
discuss this point here because the focus of this paper is on SRNs.

\begin{figure}[htbp] 
\centerline{\psfig{figure=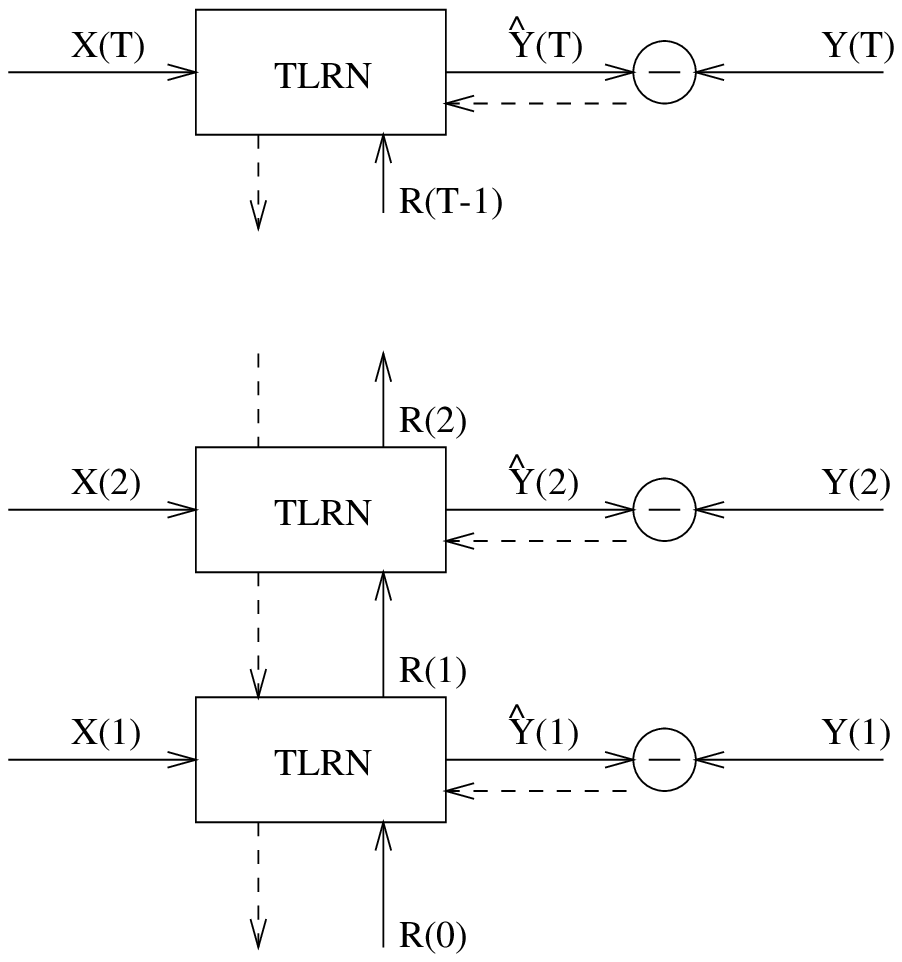,height=4.0in}}
\caption{BTT for TLRN}
\label{channel}
\end{figure}

\begin{figure}[htbp]  
\centerline{\psfig{figure=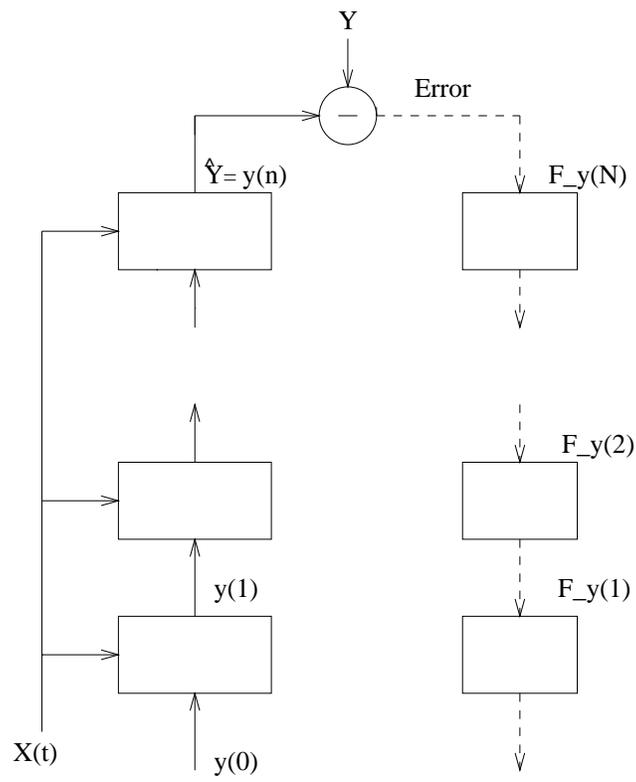,height=4.0in}}
\caption{BTT for SRN}
\label{channel}
\end{figure}

Figure 10 shows the application of BTT to training an SRN. This
figure also provides some explanation of our computer code in the
appendix. In this figure, the left-hand side(the solid arrows)
represents the neural network which predicts our desired output $Y$. (In
our example, $Y$ represents the true values of the $J$ function across
all points in the maze). Each box on the left represents a call to a feed-forward
system. The vector $X(t)$ represents the external inputs to the entire
system. In our case, $X(t)$ consists of two variables, indicating
which squares in the maze contain obstacles and which contains the goal
respectively. For simplicity, we selected the initial vector $y(0)$ as a
constant vector as we will describe below. Each call to the
feed-forward system includes calls to a subroutine which implements the
generalized MLP.

On the right-hand side of Figure 10, we illustrate the
backpropagation calculation used to calculate the derivatives. For the
SRN, unlike the TLRN, the final error depends directly only on the
output of the last iteration. Therefore the last iteration receives
feedback only from the final error but the other iterations receive
feedback only from the iterations just after them. Each box on the
right-hand side represents a backpropagation calculation through the
feed-forward system on its left. The actual backpropagation
calculation involves multiple calls to the dual subroutine $F\_net2$, which
is similar to a subroutine in chapter 8 of [2].

Notice that the derivative calculation here costs about the same amount
as the forward calculation on the left-hand side. Thus BTT is very
inexpensive in terms of computer time. However, the backpropagation
calculations do require the storage of many intermediate results. Also
we know that the human brain does not perform such extended
calculations backward through time. Therefore BTT is not a plausible
model of true brain-like intelligence. We use it here because it is
exact and therefore has the best chance to solve this difficult
problem never before solved. In future research, we may try to see whether
this problem can also be solved in a more brain-like fashion.

\subsubsection{Truncation for SRNs}

Truncation is probably the most popular method to train SRNs even
though the term truncation is not often used. For example, the "simple
recurrent networks" used in psychology are typically just SRNs adapted by
truncation[19].

Strictly speaking there are two kinds of truncation --- ordinary
one-step truncation(Figure 11) and multi-step truncation which is actually
a form of BTT. Ordinary truncation is by far the most popular. In the
derivative calculation of 
ordinary truncation, the memory inputs to the last iteration are
treated as if they were
fixed external inputs to the network. In truncation there is only one
pass of ordinary backpropagation involving only the last iteration of the
network. Many people have adapted recurrent networks in this simple way
because it seems so obvious. However, the derivatives calculated in this
way are not exactly the same because they do not totally represent the
impact of changing the weights on the final error. The reason for this is that changing the
weights will change the inputs to the final iteration. It is not right to
treat these inputs as constants because they are changed when the weights
are changed.

\begin{figure}[htbp] 
\centerline{\psfig{figure=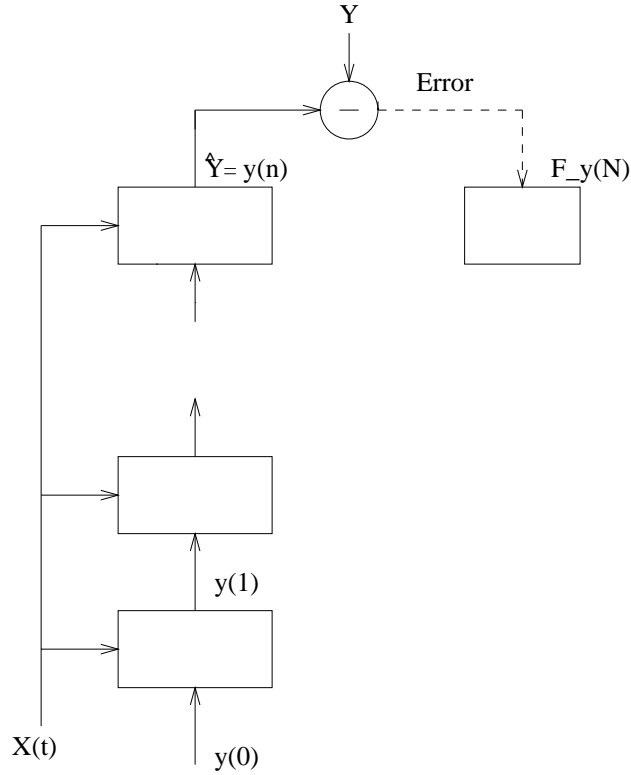,height=4.0in}}
\caption{Truncation}
\label{channel}
\end{figure}

The difference between truncation and BTT can be seen even in a simple
scalar example, where n=2 and the feed-forward calculation is linear. In
this case, the feed-forward calculation is:

\begin{equation}
y(1)=A*y(0)+B*X
\end{equation}

\begin{equation}
y(2)=A*y(1)+B*X
\end{equation}

In additon,

\begin{equation}
E=Error={\frac{1}{2}}{(Y-y(2))}^{2}
\end{equation}

\begin{equation}
\frac{\partial E}{\partial y(2)}=y(2)-Y
\end{equation}

In truncation, we use Equation (8) and deduce:

\begin{equation}
\frac{\partial E}{\partial B}={\frac{\partial E}{\partial y(2)}}*{\frac{\partial y(2)}{\partial B}}=(y(2)-Y)*X
\end{equation}

But for a complete calculation, we substitute (7) into (8), deriving:

\begin{equation}
y(2)=A^{2}*y(0)+A*B*X +B*X
\end{equation}

which yields:

\begin{equation}
\frac{\partial E}{\partial B}=(y(2)-Y)*(A*X+X)
\end{equation}

The result in Equation(11) is usually different from the result in
Equation(13), which is the true result, and comes from
BTT. Depending on the value of $A$, these results could even have
opposite signs.

In this paper, we have tried to used truncation because it is so easy
and so popular. If truncation had worked, it would be the easiest way
to solve this problem. However, it did not work.

\subsubsection{Simultaneous Backpropagation}

Simultaneous backpropagation is a method developed independently in
different forms by Werbos, Almeida and Pineda[20][21][22]. The most
general form of this method for SRNs can be found in chapter 3 of[9]
and in [23]. This method is guaranteed to converge to the
exact derivatives for the neighborhood of the equilibrium $y({\infty})$
in the case where the forward calculations have reached
equilibrium[20].

As with BTT, the derivative calculations are not expensive. Unlike BTT
there is no need for intermediate storage or for calculation backward
through time. Therefore simultaneous backpropagation could be
plausible as a model of  learning in the brain. On the other hand,
these derivative calculations do not account for the details of what
happened in the early iterations. Unlike BTT, they are not guaranteed
to be exact in the case where the final $y(n)$ is not an exact
equilibrium. Even in modeling the brain there may be some need to
train SRNs so as to improve the calculation in early iterations. In
summary, though simultaneous backpropagation may be powerful enough to
solve this problem, there was sufficient doubt that we decided to wait
until later before experimenting with this method.

\subsubsection{Error Critic}

The Error Critic, like simultaneous backpropagation, provides
approximate derivatives. Unlike simultaneous backpropagation, it has
no guarantee of yielding exact results in equilibrium. On the other
hand, because it approximates BTT directly in a statistically consistent
manner, it can account for the early iterations. Chapter 13 of [9] has
argued that the Error Critic is the only plausible model for how the
human brain adapts the TLRNs in the neocortex. It would be straightforward in principle to apply the Error Critic to training SRNs as well.

Figure 7 shows the idea of an Error Critic for TLRNs. This figure should be
compared with Figure 10. The dashed input coming into the TLRN in
Figure 7 is intended to be an approximation of the same dashed line coming
into the TLRN in the Figure 9. In effect, the Error Critic is simply
a neural network trained to approximate the complex calculations which
lead up to that dashed line in the Figure 8. The line which ends as the
dashed line in Figure 7 begins as a solid line because those derivatives
are estimated as the ordinary output of a neural network, the Error Critic. In
order to train the Error Critic to output such approximations, we use
the error calculation illustrated on the lower right of Figure 7. In this
case, the output of the Error Critic from the previous time period is
compared against a set of targets coming from the TLRN. These targets are
simply the derivatives which come out of the TLRN after one pass of
backpropagation starting from these estimated derivatives from the
later time period. This kind of training may seem a bit circular but
in fact it has an exact parallel to the kind of bootstrapping used in
the well known designs for adaptive critics or ADP.

As with simultaneous backpropagation, we intend to explore this kind
of design in the future, now that we have shown how SRNs can in fact solve
the maze problem.

\subsubsection{Forward Propagation}

The major characteristics of this method have been described
above. This method has been independently rediscovered many times with
minor variations. For
example, in 1981 Werbos called it conventional perturbation[2]. Williams
has called it the Williams -- Zipser method[5]. Narendra
has called it dynamic backpropagation.

Nevertheless, because this method is more expensive than BTT, has no
performance advantage over BTT, and is not plausible as a model of
learning in the brain, we see no reason to use this method.

\section{Two Test Problems}

In this paper we use two examples to show that the SRN design has
more general function approximation capabilities than does the MLP. Our
primary focus was on the maze problem because of its relation to 
intelligent control as discussed in Chapters 1 and 2. However, before
studying this more specialized example, we performed a few experiments
on a more general problem which we call Net A/Net B. This chapter will
discuss these two problems in more detail. 

\subsection{Net A/Net B}

In the Net A/Net B problem, our fundamental goal is to explore the
idea that the functions that an MLP can approximate are a subset of
what an
SRN can. In other words, we hypothesize that an SRN can learn to
approximate any functions which an MLP can represent without adding too
much complexity, but not vice versa. To consider the functions which
an MLP can represent, we can simply sample a set of randomly selected
MLPs, and then test the ability of SRNs to learn
those functions. Similarly we can generate SRNs at random and test the
ability of MLPs to learn to approximate the SRNs.

In order to implement this idea, we used the approach shown in
Figure 12. The first step in the process was to pick Net A at
random. In some experiments, Net A was an SRN, while in the other
experiments, it was an MLP. In all these experiments, Net B was chosen to
be the opposite kind of network from Net A. In picking Net A, we
always used the same feed-forward structure. But we used a random number
generator to set the weights. After Net A was chosen and Net B was
initialized, we generated a stream of random inputs between -1 and +1
following a uniform distribution. For each set of inputs, we trained
Net B  to try to imitate the output of Net A. Of course Net A was
fixed. The results gave an indication of the ability of Net B to
approximate Net A.

\begin{figure}[htbp] 
\centerline{\psfig{figure=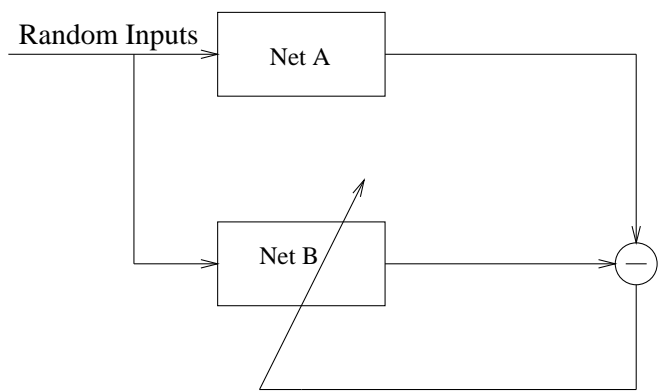,height=2.0in}}
\caption{Net A/Net B}
\label{channel}
\end{figure}

Our preliminary experiments did show that the SRNs have some advantage
over the MLPs. However, in all of these experiments, the SRN was trained
with truncation, not BTT. To fully explore all the theoretical issues
would require a much larger set of computer runs. Still, these initial
experiments were very useful in testing some general computer code
in order to prepare for the complexities of the maze problem.

\subsection{The Maze Problem}

In the classic form of the maze problem, a little robot is asked to
find the shortest path from the starting position to a goal position
on a two-dimensional surface where there are some obstacles. For
simplicity, this surface is usually represented as a kind of chess
board or grid of squares in which every square is either clear or
blocked by an obstacle. In formal terms, this means that we can
describe the state of the maze by providing three pieces of
information:
\\
(1) An array $A[ix][iy]$ which has the value 0 when the square is clear
and 1 when it is covered by an obstacle;\\
(2) The coordinates of the goal;\\
(3) The coordinates of the starting square.\\

In our case, we used a large number to represent the obstacles.

As discussed in the introduction, many researchers have trained neural
networks to learn an individual maze[5]. Our goal was to train a
network to input the array $A$ and to output $J[ix][iy]$ for all
the clear squares. According to dynamic programming, the best strategy
of motion for a robot is simply to move to that neighboring square
which has the smallest $J$.

This more general problem has not been solved before with neural
networks. For example, Houillon etc[24] initially attempted to solve
this problem with MLPs, but were unsuccessful. Widrow in several
plenary talks has reported that his neural truck backer upper has
some ability to see and avoid obstacles. However, this ability was
based on an externally developed potential function which was not
itself learned by neural networks. Such potential functions are
analogous to the $J$ function which we are trying to learn.

In fact, this maze problem can always be solved directly and
economically by dynamic programming. Why then do we bother to use a
neural network on this problem? The reason for using this test is not
because this simple maze is important for its own sake, but because
this is a very difficult problem for a learning system, and because
the availability of the correct solution is very useful for
testing. It is one thing for a human being to know the answer to a
problem. It is a different thing to build a learning system which can
figure out the answer for itself. Once the simple maze problem is
fully conquered, we can then move on to solve more difficult
navigation problems which are too complex for exact dynamic programming.

In order to represent the maze problem as a problem for supervised
learning, we need to generate both the inputs to the network(the
array $A$) and the desired outputs(the array $B$)(Refer to the Appendix). For this basic experiment,
we chose to study the example maze shown in Figure 13. In this
figure, $G$
represents the goal position, which is assigned a value of "1"; the
other numbers represent the true values of the $J$ function as calculated
by dynamic programming (subroutine "Synthesis" in the attached code in the
appendix). Intuitively each $J$ value represents the length of the
shortest path from that square to the goal. 

\begin{figure}[htbp]
\centerline{\psfig{figure=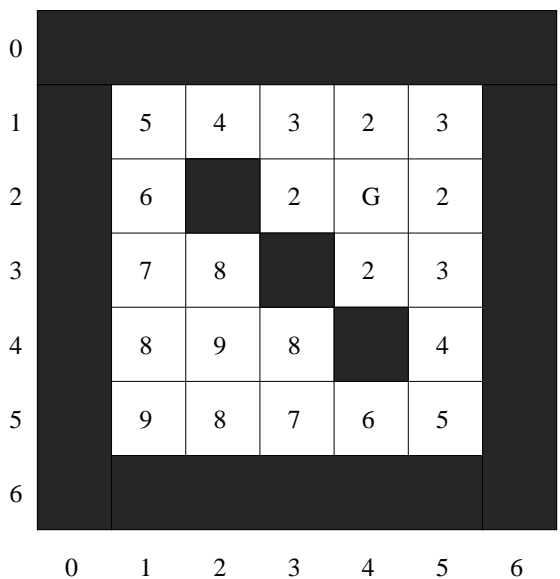,height=3.0in}}
\caption{Desired J function of a maze}
\label{channel}
\end{figure}

Initially we chose to study this particular maze because it poses some
very unique difficulties. In particular there are four equally good
directions starting from one of these squares in this maze --- a
feature which can be very confusing to neural networks, even human. If we
had used a fully connected conventional neural network, then the use of
a single test maze would have led to over-training and meaningless
results. However, as we will discuss later in this chapter, we constrained
all of our networks to prevent this problem. Nevertheless, a major
goal of our future research will be to test the ability of SRNs to
predict new mazes after training on different mazes.

This problem of maze navigation has some similarity to the problem of
connectedness described by Minsky[25]. Logically we know that the
desired output in any square can depend on the situation in any other
square. Therefore, it is hard to believe that a simple feed-forward
calculation can solve this kind of problem. On the other hand, the
Bellman equation(Equation(1)) itself is a simple recurrent equation based on
relationships between "neighboring"(successive) states. Therefore it is
natural to expect that a recurrent structure could approximate a $J$
function. The empirical results in this paper confirm these
expections.

\section{Details of Architecture and Learning Procedure}

The architecture and learning used for the net A/Net B problem were
all very standard. They will be discussed briefly in section 5.1. The
bulk of this chapter will then describe the two special feature --
cellular architecture and adaptive learning rate(ALR) used for the
maze problem.

\subsection{Details for the Net A/Net B Problem}

In all these experiments, the MLP network and the feed-forward network
$f$ in the SRN was a standard MLP with two hidden layers. The input
vector $X$ consisted of six numbers between -1 and +1. The two hidden
layers and the output layers all had three neurons. The initial
weights were chosen at random according to a uniform distribution
between -1 and +1. Training was done by standard backpropagation with
a learning rate of $0.1$.

\subsection{Weight-sharing and Cellular Architecture}

\subsubsection{What is Weight-sharing?}

In theoretical terms, weight-sharing is a generalized technique for
exploiting prior knowledge about some symmetry in the function to be
approximated. Weight-sharing has sometimes been called "windowing" or
"Lie Group" techniques.

Weight-sharing has been used almost exclusively for applications like
character recognition or image processing where the inputs form a
two-dimensional array of pixels[3][26]. In our maze problem the
inputs and outputs also form arrays of pixels. Weight-sharing leads to
a reduction in the number of weights. Fewer weights lead in turn to
better generalization and easier learning.

As an example, suppose that we have an array of hidden neurons with
voltages $net[ix][iy]$, while the input pixels form an array
$X[ix][iy]$. In that case, the voltages for a conventional MLP would be
determined by the equation:

\begin{equation}
net[i][j]={\sum_{ix,iy} W(i,j,ix,iy)*X(ix,iy)}
\end{equation}

Thus if each array has a size $20*20$, the weights form an array of size
$20*20*20*20$. This means 160,000 weights --- a very big problem. In
basic weight-sharing, this equation would be replaced by:

\begin{equation}
net[i][j]={\sum_{d1,d2} W(d1,d2)*X(i+d1,i+d2)}
\end{equation}

Furthermore, if $d1$ and $d2$ are limited to an range like $[-5, 5]$, then the
number of weights can be reduced to just over 100. Actually this would
make it possible to add two or three additional types of hidden
neurons without exceeding 1,000 weights. This trick was used by
Guyon etc[3]. They used it to develop the most successful zip code
digit recognizer in existence.

Intuitively ${AT\&T}$ justified this idea by arguing that similar patterns
in different locations have similar meanings. However, there is a more
rigorous mathematical justification for this procedure as we will see.

\subsubsection{Lie Group Symmetry and Weight-sharing}

The technique of weight-sharing in neural networks is really just a
special case of the Lie-group method pioneered much earlier by Laveen
Kanal and others in image processing. Formally speaking, if we know
that the function $F$ to be approximated must obey a certain symmetry
requirement then we can impose the same symmetry on the neural network
which we use to approximate $F$. More preciously, if $Y=F(x)$ always
implies that $MY=F(Mx)$, where $M$ is some kind of simple linear
transformation, then we can require that the neural network possess the
same symmetry.

Both in image processing and in the maze problem, we can use the
symmetry with respect to those transformations $M$ which move all the pixels by the same
distance to the left, to the right or up and down. In the language of
physics, these are called spatial translations.

Because we know that the best form of the neural network must also
obey this symmetry, we have nothing to lose by restricting our weights
as required by the symmetry.

\subsubsection{How We implemented Weight-sharing}

In order to exploit Lie group symmetry in a rigorous way, we first
reformulated the task to be solved so as to ensure exact Lie group
symmetry. To do this, we designed our neural network to solve the
problem of maze defined over a torus. For our purposes, a torus was
simply an $N$ by $N$ square where the right-hand neighbor of $[i,N]$ is the point $[i,0]$, and likewise
for the other edges. This system can still solve an ordinary maze as
in Figure 13, where the maze is surrounded by walls of obstacles.

Next we used a cellular structure for our neural network including
both the MLPs and SRNs. A cellular structure means that the network is
made up of a set of cells each made up of a set of neurons. There is
one cell for each square in the maze. The neurons and the weights in
each cell are the same as those in any other cell. Only the inputs and
outputs are different because they come from different locations.

The general idea of our design is shown in Figure 14. Notice
that each cell is made up of two parts: a connector part and a local
memory part which includes 4 neighbors and the memory from itself in
Fingure 15. The connector part receives the inputs to the cell and
transmits its output to all four neighboring cells.  In addition, the local memory part sends all its
outputs back as inputs, but only to the same cell. Finally the
forecast of $J$ is based on the output of the local memory part.

\begin{figure}[htbp] 
\centerline{\psfig{figure=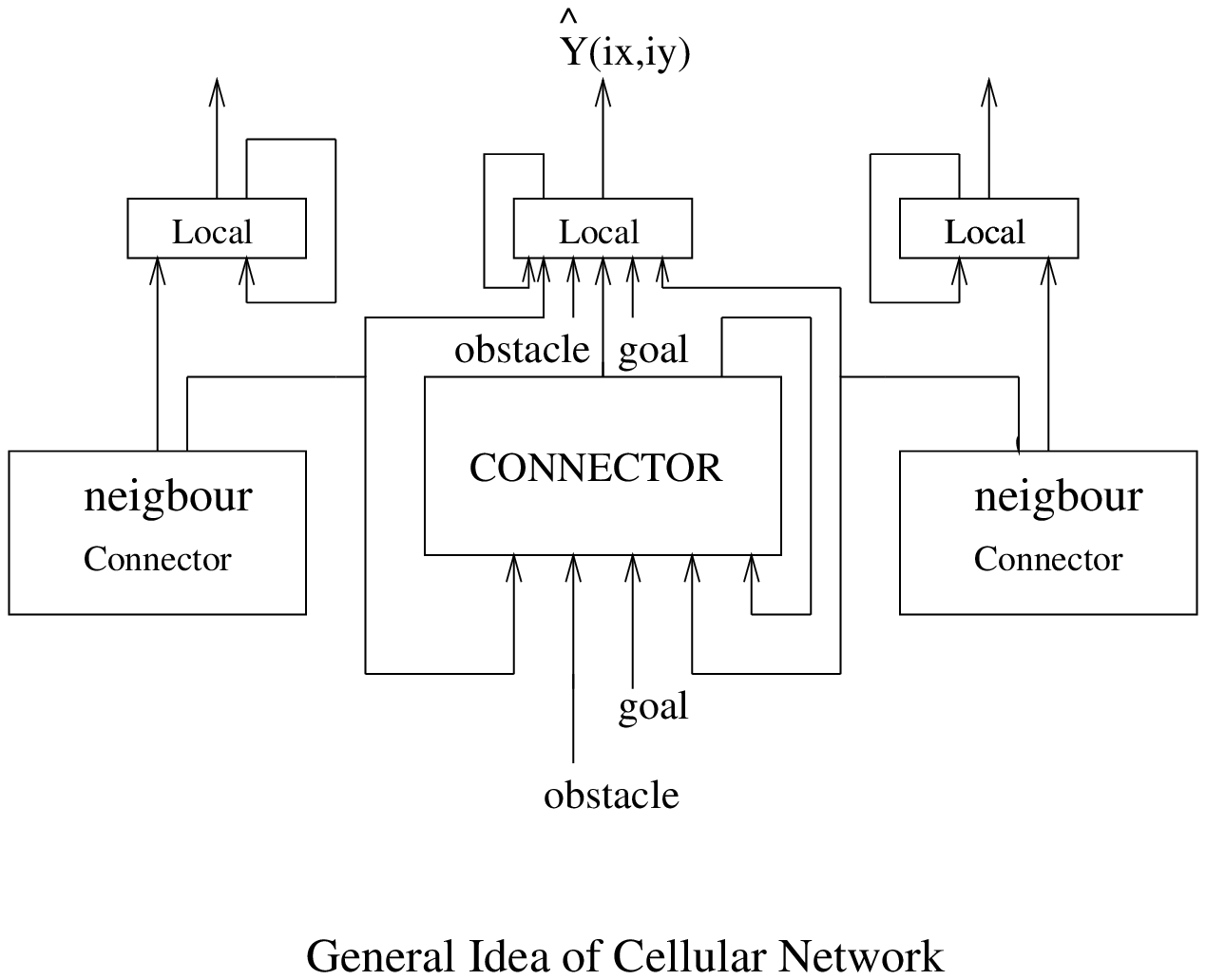,height=3.0in}}
\caption{General Idea of the Cellular Network}
\label{channel}
\end{figure}

\begin{figure}[htbp] 
\centerline{\psfig{figure=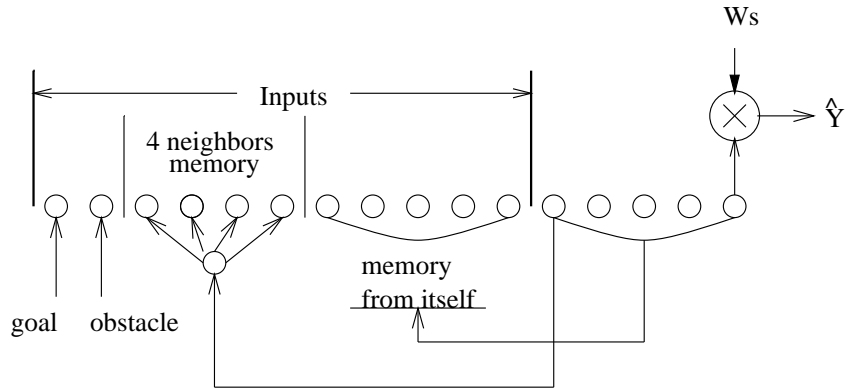,height=2.0in}}
\caption{Inputs, Outputs and Memory of Each Cell}
\label{channel}
\end{figure}

The exact structure which we used is shown completely in Figure 15. In
this figure it can be seen that each cell receives 11 inputs on each
iteration. Two of these inputs represent the goal and obstacle
variables, $A[ix][iy]$ and $B[ix][iy]$, for the current pixel. The next
four inputs represent the outputs of the connector neuron from the
four neighboring cells from the previous iteration. The final five
inputs are simply the outputs of the same cell from the previous
iteration. Then after the inputs, there are only five actual
neurons. The connector part is only one neuron in our case. The local
memory part is four neurons. The prediction of $J[ix][iy]$ results from
multiplying the output of the last neuron by $Ws$, a weight used to
rescale the output.

To complete this description, we must specify how the five active
neurons work. In this case, each neuron takes inputs from all of the
neurons to its left, as in the generalized MLP design[2]. Except
for $\hat{J}$, all of the inputs and outputs range between -1 and 1,
and the
tanh function is used in place of the usual sigmoid function.

To initialize the SRN on iteration zero, we simply picked a reasonable
looking constant vector for the first four neurons out of the five. We
set the initial starting value to -1. For the last neuron, we set it to
0. In future work, we shall probably experiment with the adaptation of
the starting vector $y(0)$.

In order to backpropagate through this entire cellular structure, we
simply applied the chain rule for ordered derivatives as described in
[2].

\subsection{Adaptive Learning Rate}

In our initial experiments with this structure, we used ordinary
dynamic programming with only one special trick. The trick was that we
set the number of iterations for SRN to only 1 on the first 20
trials, and then to 2 for the next 20 trials... and so on up until
there were 20 iterations.

We found that ordinary weight adjustment led to extremely slow
learning due to oscillation. This was not totally unexpected because
slow learning and oscillation are a common result of simple steepest
descent methods. There are many methods available to accelerate the
learning. Some of these like the DEKF method developed by Ford Motor
Company are similar to quasi-Newton methods[27] which are very powerful but
also somewhat expensive. For this work we chose to use a method called
the adaptive learning rate(ALR) as described in chapter 3 of [9]. This
method is relatively simple and cheap, but far more flexible and
powerful than other simple alternatives.

In this method, we maintain a single adapted learning rate for each
group of weights. In this case, we chose three groups of
weights:
\\
1. The weight $Ws$ used for rescaling of the output;\\
2. The constant or bias weights $ww$;\\
3. All the other weights $W$.\\

For each group of weights the learning rate is updated on each trial
according to the following formula:

\begin{equation}
LR(t+1)=LR(t)*(0.9+0.2*{\frac{\sum_{k}W_{k}(t)*W_{k}(t-1)}{\sum_{k}W_{k}(t-1)*W_{k}(t-1)}})
\end{equation}
where the sum over $k$ actually refers to the sum over all weights in
the same group. In addition, to prevent overshoot, we would reset the
learning rate to:

\begin{equation}
\frac{LR*E} {\sum_{k}{({\frac{{\partial}E}{{\partial}W_{k}}})}^{2}}
\end{equation}
where the sum is taken over all weights. In this special case where
the error on the next iteration would be predicted to be less than
zero, i.e.:

\begin{eqnarray}
& &E-\sum_{k}({W_{k}}(t+1)-W_{k}(t))*{\frac{{\partial}E}{{\partial}W_{k}}}(t)\nonumber
\\
&=&E-\sum_{k}(LR*{\frac{{\partial}E}{{\partial}W_{k}}}(t))*{\frac{{\partial}E}{{\partial}W_{k}}}(t)\nonumber
\\
&=&E-LR*\sum_{k}{({\frac{{\partial}E}{{\partial}W_{k}}}(t))}^{2}<0
\end{eqnarray}
where $W_{k}(t+1)$ is the new value for the weights which would be used if
the learning rates were not reset. In our case, we modified this procedure
slightly to apply it separately to each group of weights.

After the adaptive learning rates were installed the process of
learning became far more reliable. Nevertheless, because of
the complex nature of the function $J$, there was still some degree of
local minimum problem. For our purposes, it was good enough to simply
try out a handful of initial values which we guessed at random. However,
in future research, we would like to explore the concept of shaping as
described in [9].

\section{Simulation Results and Conclusions}

In this chapter, we will see some simulation results for the two test
problems discussed before. From analyzing the results, we can
conclude that compared to the MLPs, the SRNs are more powerful in nonsmooth
function approximation. In addition, our new design --- the cellular
structure --- can really solve the maze problem.

\subsection{Results for the Net A/Net B Problem}

From Figure 16 to Figure 19 we can see that the SRN using the same
three-layered neural network structure(9 inputs, 3 outputs, and 3
neurons for each hidden layer) as the MLP can achieve better simulation
result. The SRN not only converged more rapidly than the MLP(Figure 16 and
Figure 17, but also
reached a smaller error(Figure 18 and Figure 19), about $1.25*10^{-4}$,
while the MLP reached $5*10^{-4}$. Thus we can say that, in this
typical case, an SRN has better ability to learn an MLP than an MLP to
learn an SRN.

\begin{figure}[htbp] 
\centerline{\psfig{figure=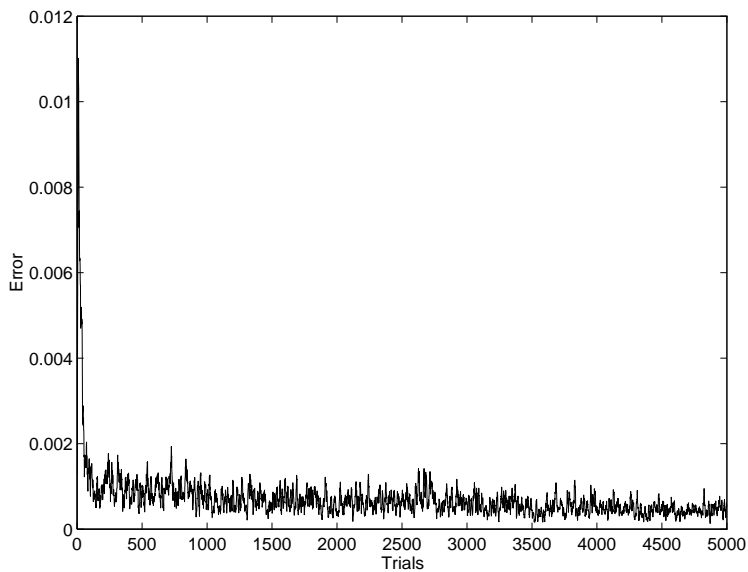,height=3.0in}}
\caption{The MLP learned the SRN}
\label{channel}
\end{figure}

\begin{figure}[htbp] 
\centerline{\psfig{figure=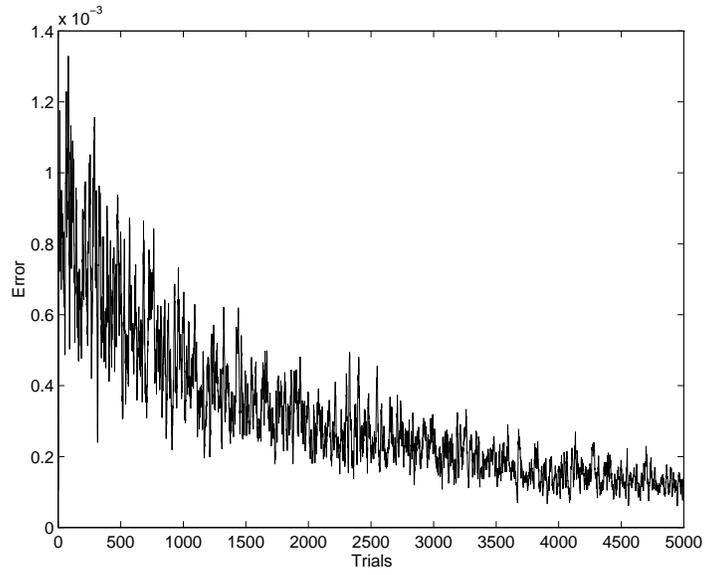,height=3.0in}}
\caption{The SRN learned the MLP}
\label{channel}
\end{figure}

\begin{figure}[htbp] 
\centerline{\psfig{figure=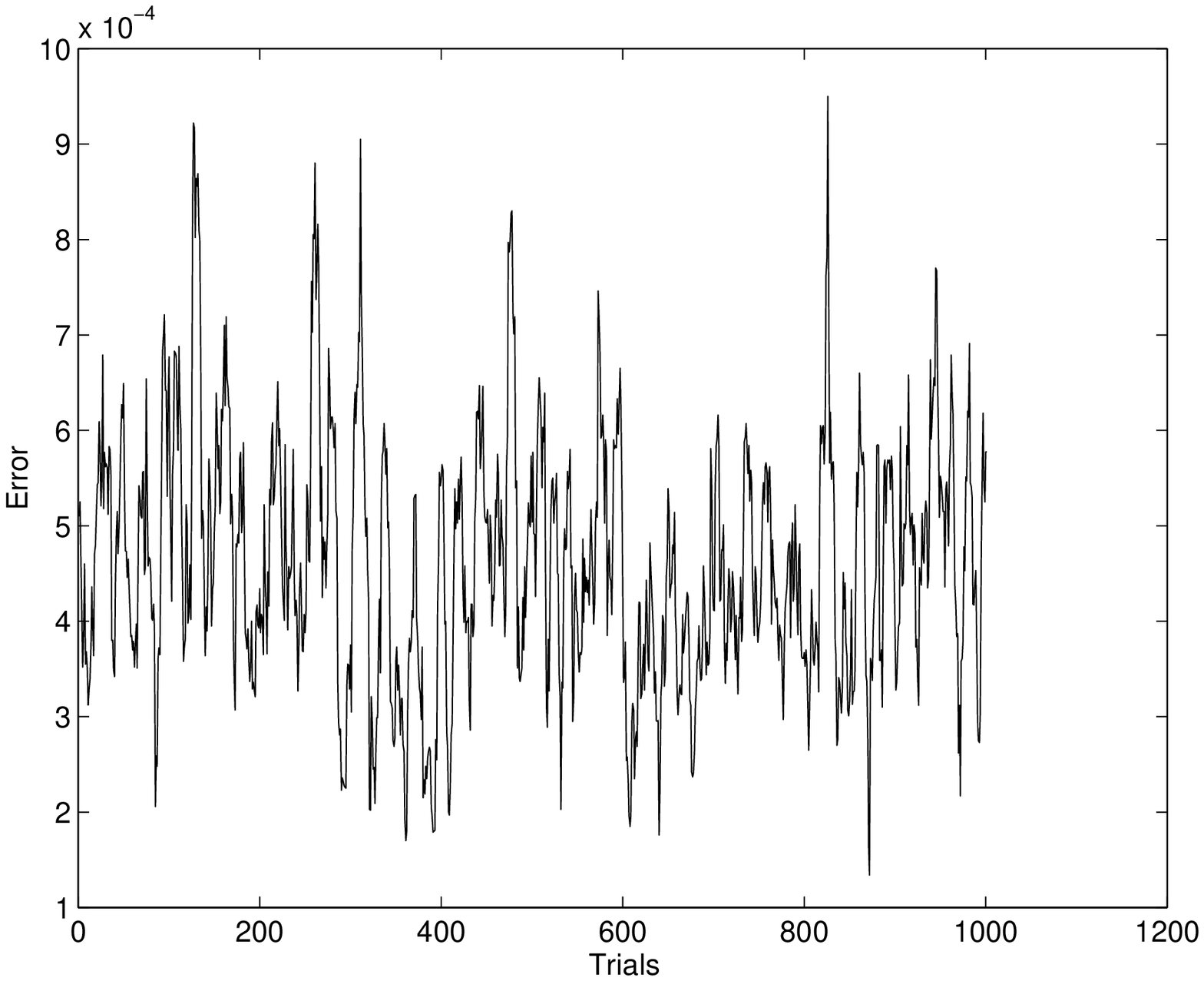,height=3.0in}}
\caption{The last 1000 trials of figure 16}
\label{channel}
\end{figure}

\begin{figure}[htbp] 
\centerline{\psfig{figure=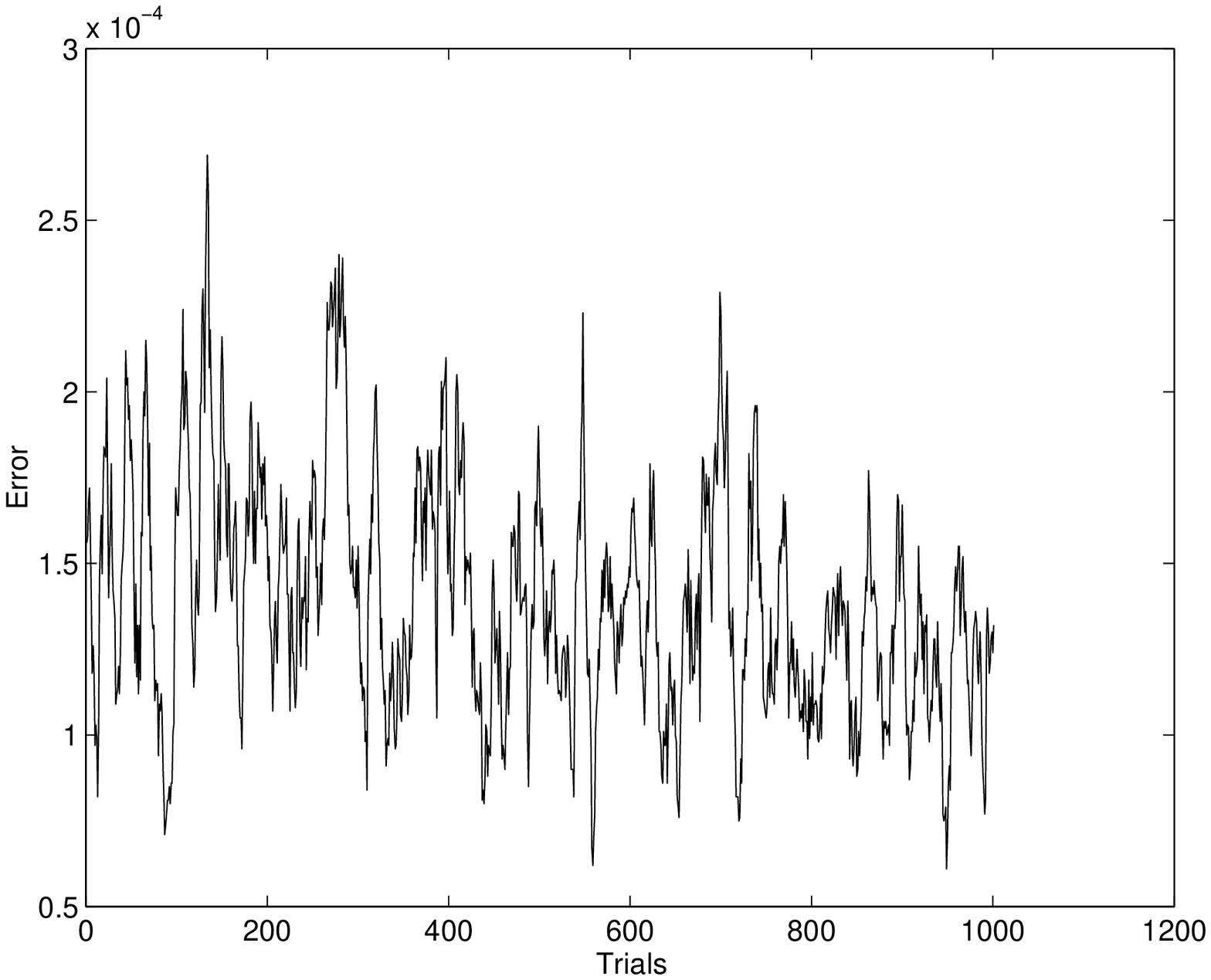,height=3.0in}}
\caption{The last 1000 trials of figure 17}
\label{channel}
\end{figure}

\subsection{Results for the Maze Problem}

There are two parts of the results for the maze problem.

First, we compare the $J$ function in each pixels of the same maze as
predicted by an SRN trained by BTT and an SRN trained by truncation
respectively with the actual J function for the maze. Figure 20 and
Figure 21
show that the SRN trained by BTT can really approximate the $J$ function,
but the SRN trained by truncation cannot. Moreover, the SRN trained by BTT can
learn the ability to find the optimal path from the start to the goal
as calculated by dynamic programming. Although there is some error in
the approximation of $J$ by the SRN trained by BTT, the errors are
small enough that a system governed by the approximation of $J$ would
always move in an optimal direction.

\begin{figure}[htbp] 
\centerline{\psfig{figure=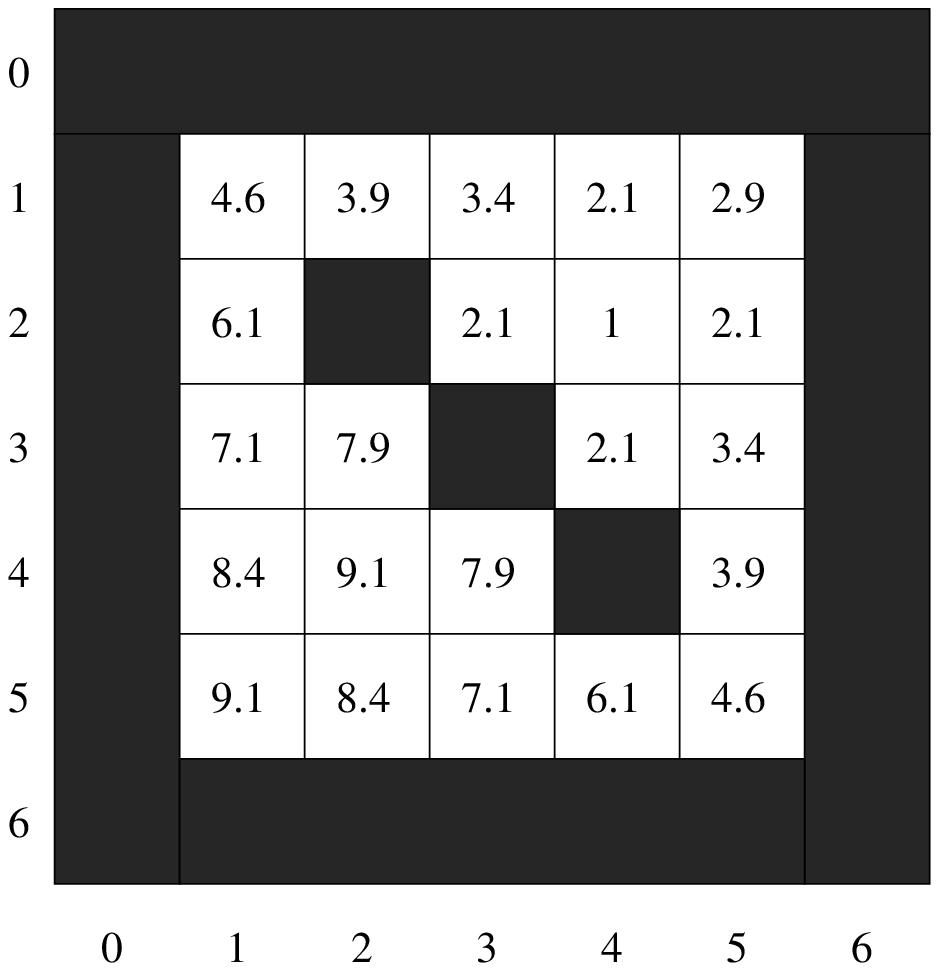,height=3.0in}}
\caption{J function as predicted by SRN-BTT(I)}
\label{channel}
\end{figure}

\begin{figure}[htbp]
\centerline{\psfig{figure=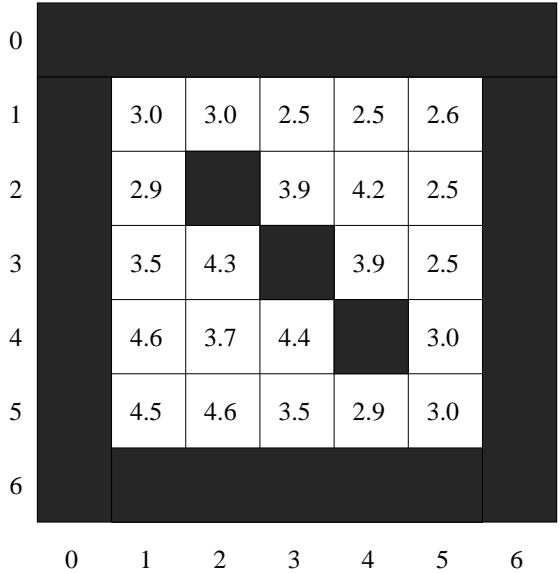,height=3.0in}}
\caption{J function as predicted by SRN-Truncation(I)}
\label{channel}
\end{figure}

Second, we show some error curves from Figure 22 to Figure 27. From the
figures we can see the error curve of SRN trained by BTT not only
converged more rapidly than the curve of the SRN trained by truncation,
but also reached a much smaller level of error. The errors with the MLP
did not improve at all after about 80 trials(Figure 26 and Figure 27).

\begin{figure}[htbp] 
\centerline{\psfig{figure=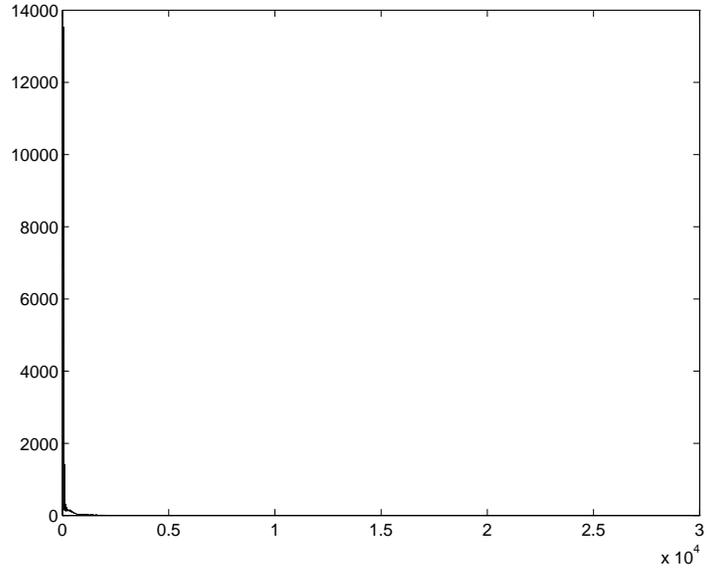,height=3.0in}}
\caption{Error curve of J function as predicted by SRN-BTT(II)}
\label{channel}
\end{figure}

\begin{figure}[htbp] 
\centerline{\psfig{figure=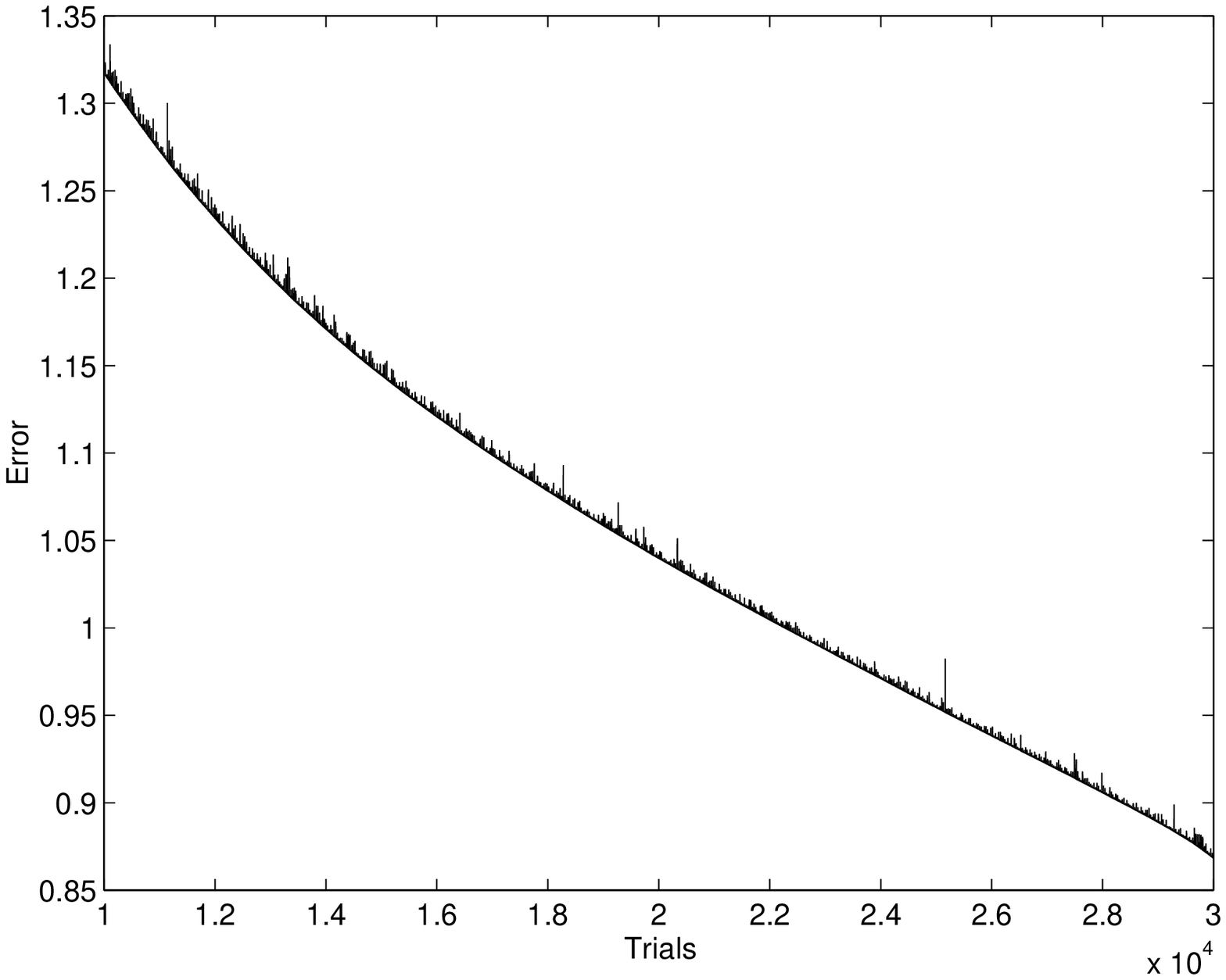,height=3.0in}}
\caption{Error curve of J function as predicted by SRN-BTT(III)}
\label{channel}
\end{figure}

\begin{figure}[htbp] 
\centerline{\psfig{figure=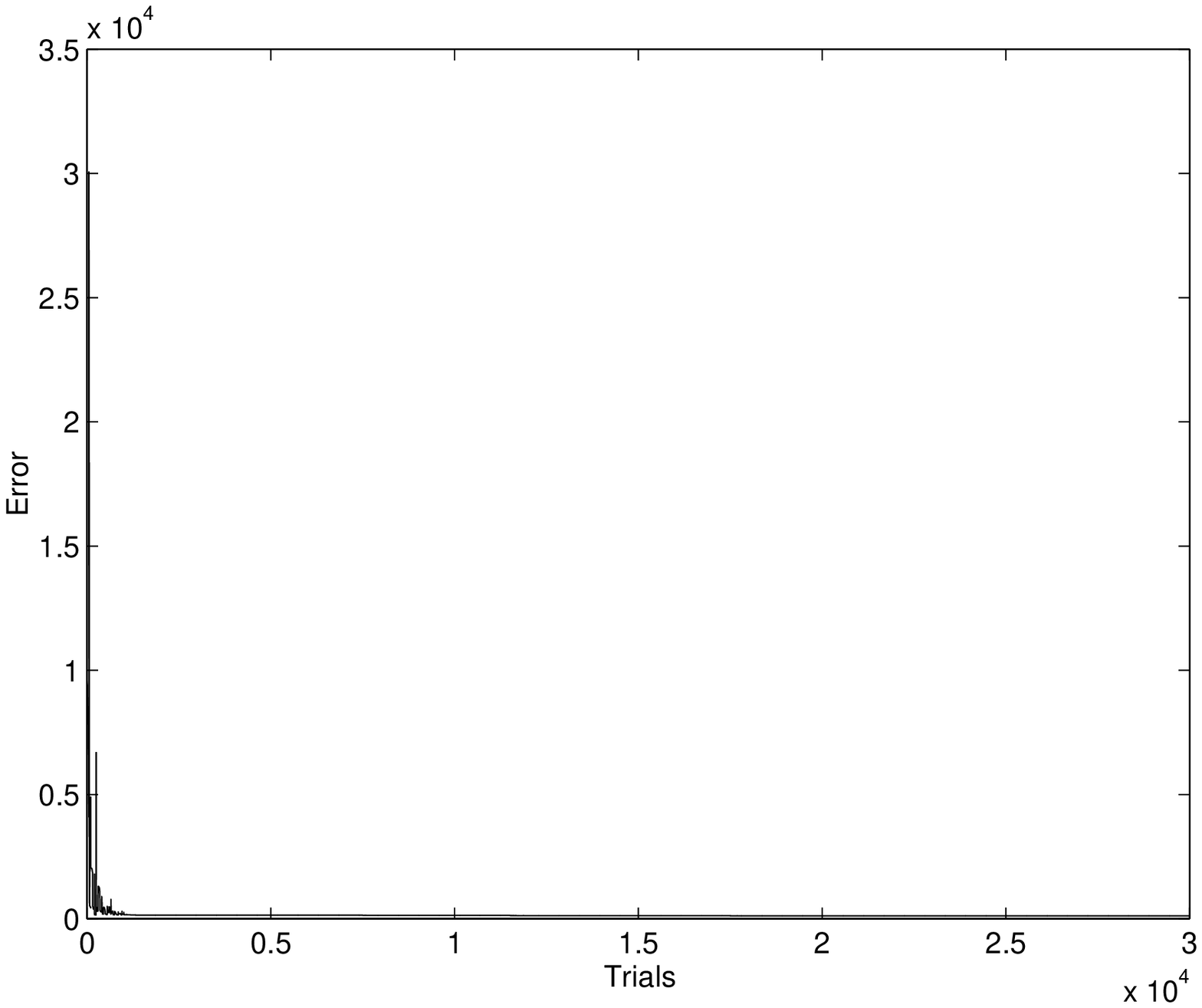,height=3.0in}}
\caption{Error curve of J function as predicted by SRN-Truncation(II)}
\label{channel}
\end{figure}

\begin{figure}[htbp]
\centerline{\psfig{figure=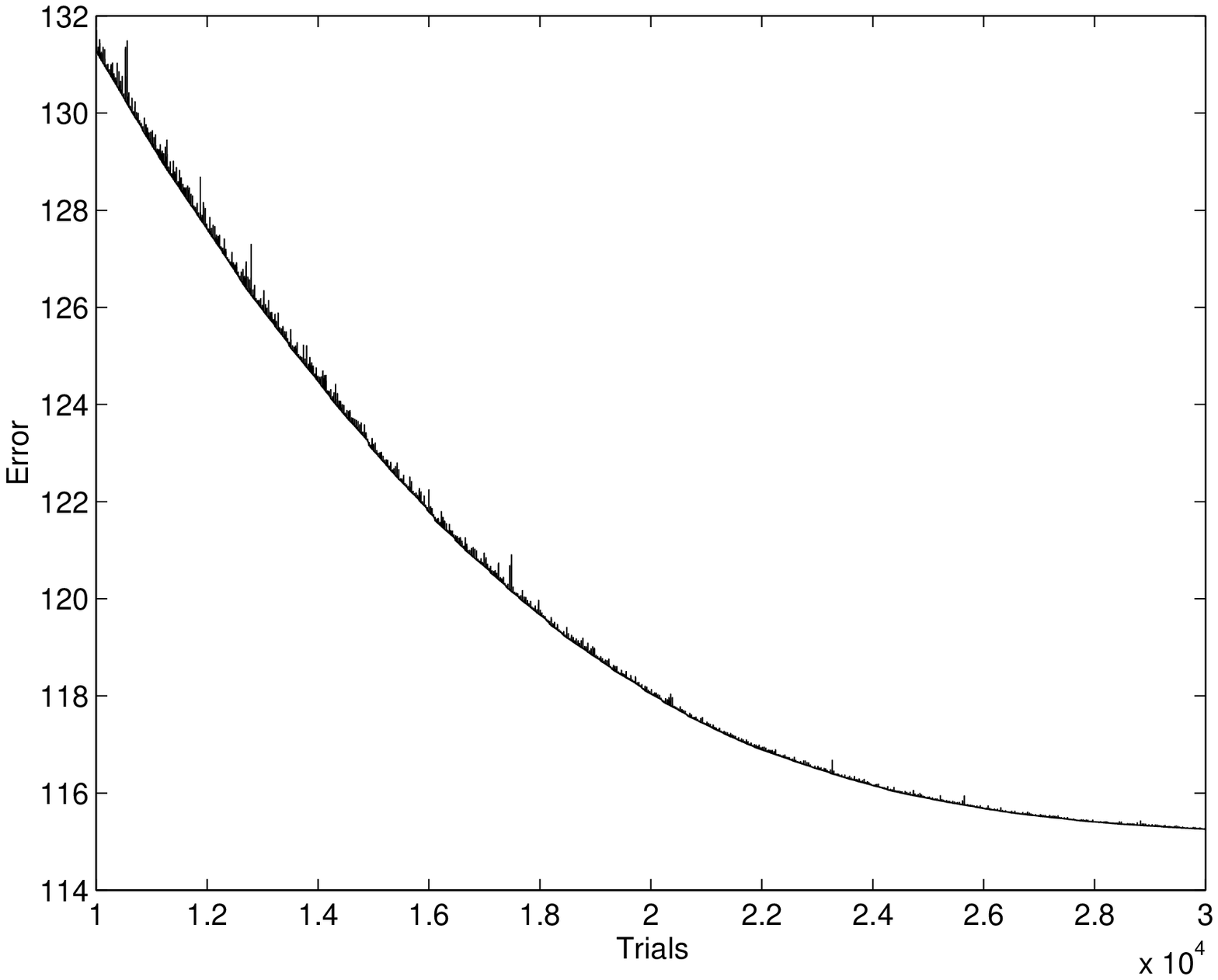,height=3.0in}}
\caption{Error curve of J function as predicted by SRN-Truncation(III)}
  \label{channel}
\end{figure}

\begin{figure}[htbp] 
\centerline{\psfig{figure=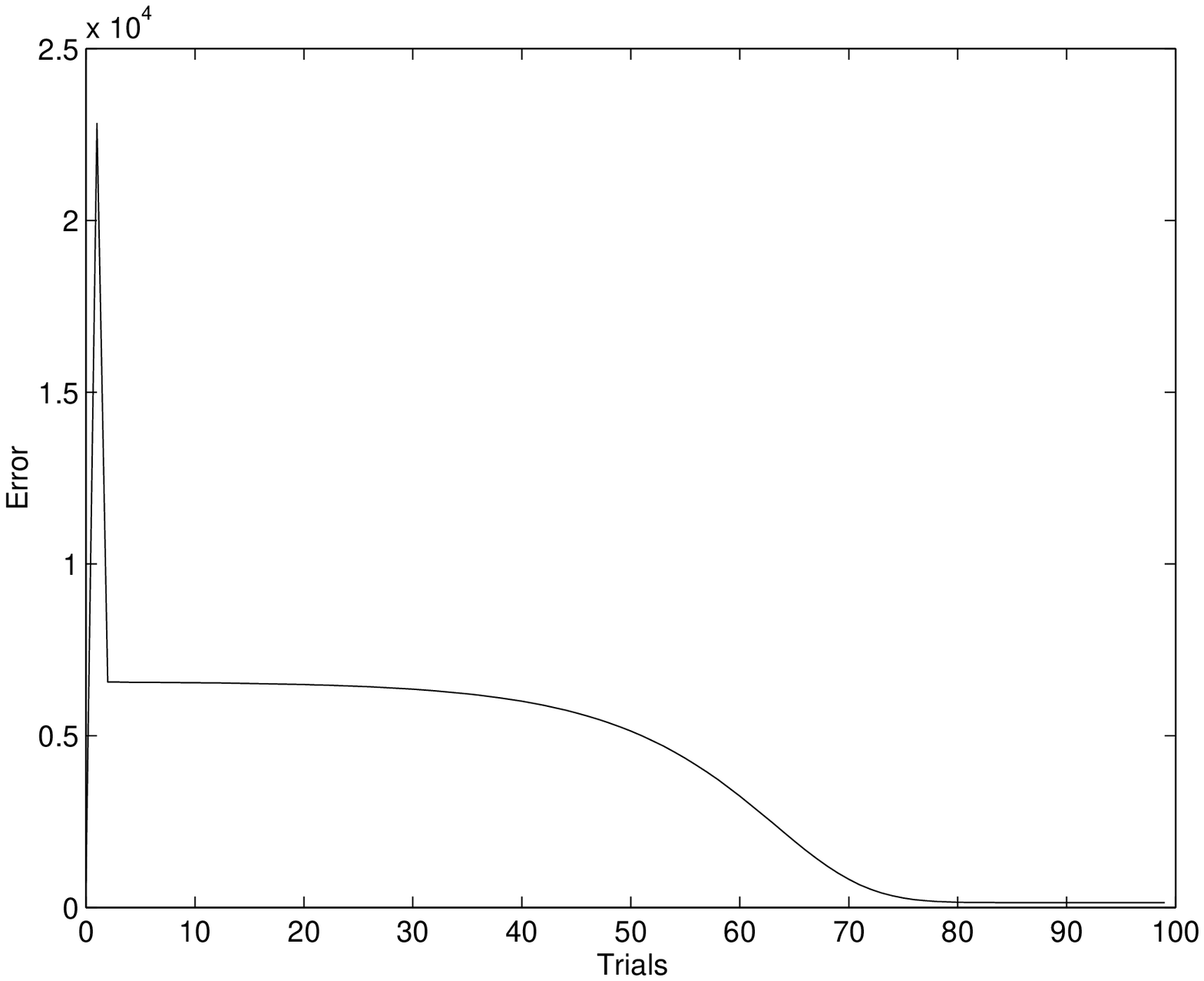,height=3.0in}}
\caption{Error curve of J function as predicted by MLP(I)}
\label{channel}
\end{figure}

\begin{figure}[htbp] 
\centerline{\psfig{figure=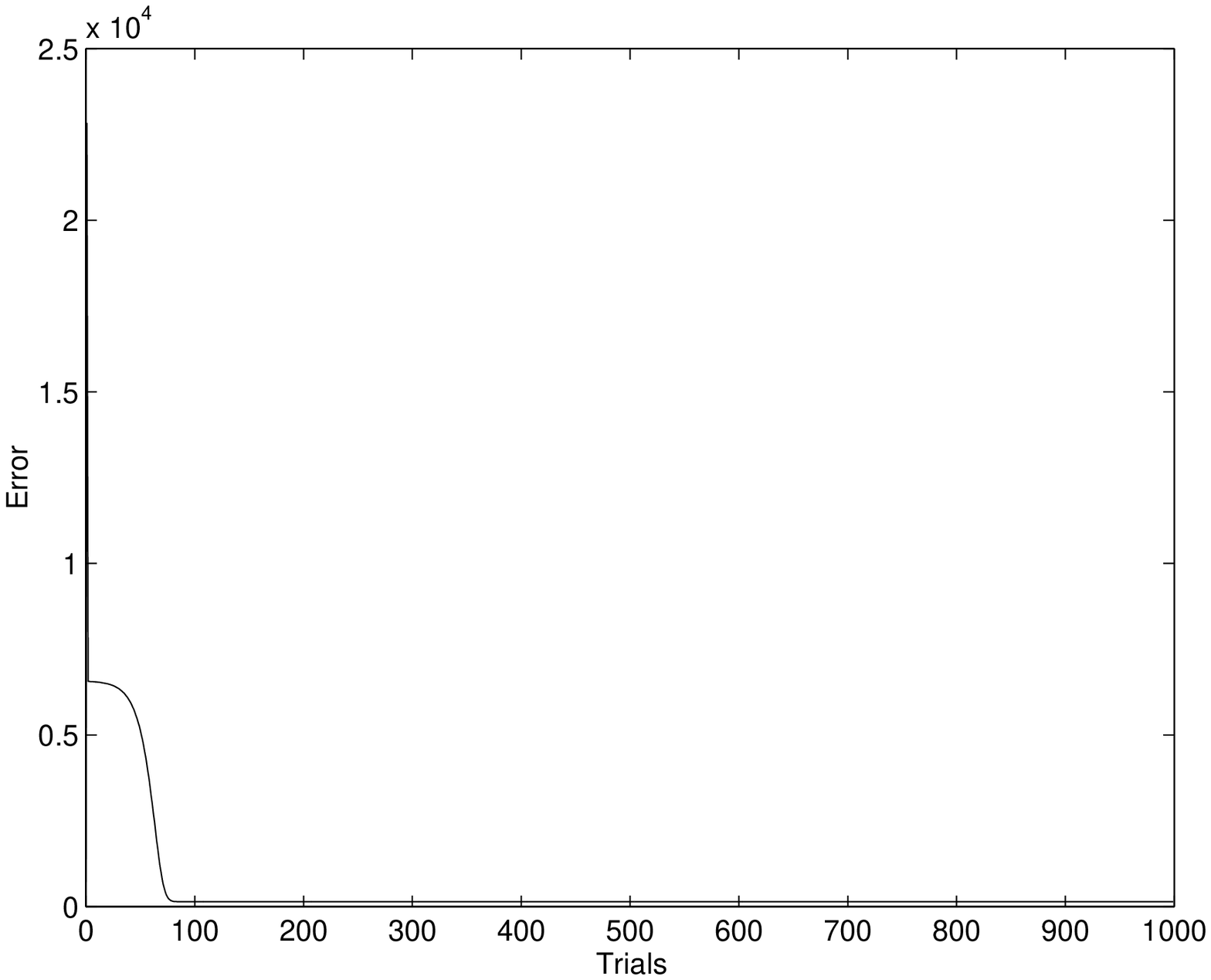,height=3.0in}}
\caption{Error curve J function as predicted by MLP(II)}
\label{channel}
\end{figure}

\subsection{Conclusions} 
In this paper, we have described a new neural network design for $J$
function approximation in dynamic programming. We have tested this
design in two test problems: Net A/Net B and the maze problem. In the
Net A/ Net B problem, we showed that SRNs can learn to approximate
MLPs better than MLPs can learn SRNs.

In the maze problem, a much more complex problem, we showed that we
can achieve good results only by training an SRN with a combination of
BTT and adaptive learning rates. In addition, we needed to use a
special design --- a cellular structure --- to solve this problem. On
the other hand, neither an MLP nor an SRN trained by truncation could
solve this problem.

Now that it has been proven that neural networks can solve these kinds
of problems, the next step in research is to consider many variations
of these problems in order to demonstrate generalization ability and
the ability to solve optimization problems while the $J$ function
is not known.

\vspace{1cm}
\begin{center}
{\large \bf Acknowledgments}
\end{center}

{The work described herein was truly collaborative work. More than half of
it was performed when both authors worked together at the University
of Maryland, College Park, or a nearby location. Neither author received any financial
support covering the time when this paper was written; however, this work
would have been impossible without prior support and ongoing connections
involving: the University of Maryland (Prof. John S. Baras and Robert
W. Newcomb); the National University of Singapore; the IEEE Singapore Section; Scientific
Cybernetics Inc.; the International Joint Conference on Neural Networks;
the National Science Foundation of the U.S. and the National Natural
Science Foundation of China.}

\appendix
\section{Appendix: The program of the maze problem using SRN trained by BTT}

\begin{verbatim}
/* The program of the maze problem using SRN trained by BTT*/
/* Using the SRN trained by BTT to learn the optimal path of a 5*5 maze:   */
/* Learning Rate Adaptive  --- Lr_Ws,Lr_W,Lr_ww */
/* When change line 136 and 137 into p=0 then the program will be MLP*/
/* When change line 243 into F_x[N+i]=0 then the program will be the
SRN trined by Truncation*/

#include <stdio.h>
#include <stdlib.h>
#include <math.h>
#include <time.h>


void F_NET2(double F_Yhat, double W[30][30],double x[30],int n,int m,int N, 
double F_W[30][30], double F_net[30],double F_Ws[30],double Ws,double F_x[30]);

void NET(double W[30][30],double x[30],double ww[30],
                int n, int m, int N, double Yhat[30]);
void synthesis(int B[30][30],int A[30][30],int n1,int n2);
void pweight(double Ws,double F_Ws_T,double ww[30],double F_net_T[30],
                double W[30][30], double F_W_T[30][30],int n,int N,int m); 

int minimum(int s,int t,int u,int v);
int min(int k,int l);
double f(double x);

void main()
{
        int i,j,it,iz,ix,iy,lt,m,maxt,n,n1,n2,nn,nm,N,p,q,po,t,TT;
        int A[30][30],B[30][30];
        double a,b,dot,e,e1,e2,es,mu,s,sum,F_Ws_T,Ws,F_Yhat,wi;

        double W[30][30],x[30],F_net_T[30],F_Ws[30],F_W_O[30][30],F_W[30][30];
        double F_W_T[30][30],F_net[30],ww[30],yy[21][12][8][8];
        double Yhat[30],F_y[21][12][8][8],F_x[30],F_Jhat[30][30];
        double S_F_W1,S_F_W2,Lr_W,S_F_net1,S_F_net2,Lr_ww,Lr_Ws,F_Ws_O;
        double y[50][50],F_net_O[50], F_Ws1, F_Ws2,W_O[50][50],ww_O[50],Ws_O;
        FILE *f;

        /* Number of inputs,neurons and output:7,3,1 */
        /* 'n' is the number of the active neurons */ 
        /* 'm' and 'N' both are the number of inputs */
        /* 'nm' is the number of memory is: 5 */
        /* 'nn+1'*'nn+1' is the size of the maze' */
        /* 'TT' is the number of trials */
        /* 'lt' is the number of the interval time */
        /* 'maxt' is the max number for T in figure[8] */
        /* Lr-Ws,Lr_ww and Lr_W are the learning rates for Ws,ww and W*/
        
        a=0.9; b=0.2;
        n=5;m=11;N=11;nn=6;nm=5;TT=30000;lt=50;maxt=20;wi=25;Ws=40;
        e=0;po=pow(2,31) -1;

        /* Initial values of Old  */
        
        F_Ws_O=1;
        for(i=m+1;i<N+n+1;i++)
        {
               for(j=1;j<i;j++)
                        F_W_O[i][j]=1;
               F_net_O[i]=1;
         }
        Lr_W=Lr_ww=Lr_Ws=10;
        
        /* Initial values of weights */
        
        for (i=1;i<N+n+1;i++)
                x[i]=0;
        for (i=m+1;i<N+n+1;i++)
                for (j=0;j<i;j++)
                {
                        srand(rand());
                        W[i][j]=rand();
                        W[i][j]=0.2091;
                }
        for (i=m+1;i<N+n+1;i++)
        {
                srand(rand());
                ww[i]=rand();
                ww[i]=0.00678;
        }

        /* Input Maze */
        
        n2=5*5;
        n1=5*5-1;
        for (i=0;i<7;i++)
                for(j=0;j<7;j++)
                {
                        if ((i==0)||(j==0)||(i==6)||(j==6))
                                B[i][j]=n2;
                        else
                                B[i][j]=n1;
                }
        /* Generate Obstacle */

        B[2][2]=B[3][3]=B[4][4]=n2;

        /* Generate Start */

        B[2][4]=1;

        for (i=0;i<7;i++)
                for(j=0;j<7;j++)
                {
                        A[i][j]=0;
                }

        /* Desired outputs */

        synthesis(B,A,n1,n2);

        if((f=fopen("results5","w"))==NULL) {
        printf("Cannot open file");
        exit(1);
        }
        
        /* Learning Pattern */
        
        for(t=0;t<TT;t++)
        {
                for(i=m+1;i<N+n+1;i++)
                {
                        for(j=1;j<i;j++)
                        {
                                F_W_T[i][j]=0;
                        }
                        F_net_T[i]=0;
                }
                for (i=1;i<n;i++)
                        for(ix=0;ix<nn+1;ix++)
                                for(iy=0;iy<nn+1;iy++)
                                        yy[0][i][ix][iy]=-1;
                for(ix=0;ix<nn+1;ix++)
                                for(iy=0;iy<nn+1;iy++)
                                        yy[0][n][ix][iy]=0;

                e=F_Ws_T=s=0;

        /* If the next two lines are changed into p=0 then it is MLP */
                p=(t/lt)+1;
                p=(p<maxt ? p:maxt);
                for(q=0;q<p+1;q++)
                {
                        e=0;
                        for(ix=0;ix<nn+1;ix++)
                                for(iy=0;iy<nn+1;iy++)
                                {
                                        if (B[ix][iy]==25)
                                                x[1]=B[ix][iy];
                                        else if (B[ix][iy]!=1)
                                                x[1]=0;
                                        x[2]=1;
                                        if (ix!=0)
                                                x[3]=yy[q][1][ix-1][iy];
                                        else
                                                x[3]=yy[q][1][nn][iy];
                                        if (iy!=0)
                                                x[4]=yy[q][1][ix][iy-1];
                                        else
                                                x[4]=yy[q][1][ix][nn];
                                        if (ix!=nn)
                                                x[5]=yy[q][1][ix+1][iy];
                                        else
                                                x[5]=yy[q][1][0][iy];
                                        if (iy!=nn)
                                                x[6]=yy[q][1][ix][iy+1];
                                        else                         
                                                x[6]=yy[q][1][ix][0];
                                        for (i=1;i<n+1;i++)
                                                x[6+i]=yy[q][i][ix][iy];
                                        NET(W,x,ww,n,m,N,Yhat);
                                        for (i=1;i<n+1;i++)
                                                yy[q+1][i][ix][iy]=Yhat[i];
                                }
                }
                e=0;
                for (ix=0;ix<nn+1;ix++)
                        for(iy=0;iy<nn+1;iy++)
                        {
                                if (t==(TT-1))
                                        y[ix][iy]=yy[p+1][n][ix][iy];
                                if (B[ix][iy]!=25)
                                        F_Jhat[ix][iy]=Ws*yy[p+1][n][ix][iy]
                                                        -A[ix][iy];
                                else
                                        F_Jhat[ix][iy]=0;
                                e+=F_Jhat[ix][iy]*F_Jhat[ix][iy];
                        }
                printf("\n t e %d %e",t,e);
                fprintf(f,"\n%d %e",t,e);  

        /* Initialize F_y */

                for(q=1;q<21;q++)
                        for(ix=0;ix<nn+1;ix++)
                                for(iy=0;iy<nn+1;iy++)
                                        for(i=1;i<n+1;i++)
                                                F_y[q][i][ix][iy]=0;
                for(q=p;q>-1;q--)
                {
                        for (ix=0;ix<nn+1;ix++)
                                for(iy=0;iy<nn+1;iy++)
                                {
                                        if (B[ix][iy]==25)
                                                x[1]=B[ix][iy];
                                        else if (B[ix][iy]!=1)
                                                x[1]=0;
                                        x[2]=1;
                                        if (ix!=0)
                                                x[3]=yy[q][1][ix-1][iy];
                                        else
                                                x[3]=yy[q][1][nn][iy];
                                        if (iy!=0)
                                                x[4]=yy[q][1][ix][iy-1];
                                        else
                                                x[4]=yy[q][1][ix][nn];
                                        if (ix!=nn)
                                                x[5]=yy[q][1][ix+1][iy];
                                        else
                                                x[5]=yy[q][1][0][iy];
                                        if (iy!=nn)
                                                x[6]=yy[q][1][ix][iy+1];
                                        else
                                                x[6]=yy[q][1][ix][0];
                                        for (i=1;i<n+1;i++)
                                                x[6+i]=yy[q][i][ix][iy];
                                        NET(W,x,ww,n,m,N,Yhat);
                                        if (q==p)
                                        {
                                                F_Yhat=F_Jhat[ix][iy];
                                                for(i=1;i<n+1;i++)
                                                        F_x[N+i]=0;
                                        }
                                        else
                                        {
                                                F_Yhat=0;
                                                for(i=1;i<n+1;i++)
                                                  F_x[N+i]=F_y[q+1][i][ix][iy];
                                        }
                                        F_NET2(F_Yhat,W,x,n,m,N,F_W,F_net,
                                                F_Ws,Ws,F_x);
                                        if (ix!=0)
                                                F_y[q][1][ix-1][iy]+=F_x[3];
                                        else
                                                F_y[q][1][nn][iy]+=F_x[3];
                                        if (iy!=0)
                                                F_y[q][1][ix][iy-1]+=F_x[4];
                                        else
                                                F_y[q][1][ix][nn]+=F_x[4];
                                        if (ix!=nn)
                                                F_y[q][1][ix+1][iy]+=F_x[5];
                                        else
                                                F_y[q][1][0][iy]+=F_x[5];
                                        if (iy!=nn)
                                                F_y[q][1][ix][iy+1]+=F_x[6];
                                        else
                                                F_y[q][1][ix][0]+=F_x[6];
                                        for(i=1;i<n+1;i++)
                                                F_y[q][i][ix][iy]+=F_x[6+i];

                                        if (q==p) F_Ws_T+=F_Ws[1];

                                        for(i=m+1;i<N+n+1;i++)
                                        {
                                                for(j=1;j<i;j++)
                                                {
                                                        F_W_T[i][j]+=F_W[i][j];
                                                }
                                                F_net_T[i]+=F_net[i];
                                        }
                                }
                }
                dot=0;
                for(i=m+1;i<N+n+1;i++)
                        for(j=1;j<i;j++)
                        {
                                dot+=F_W_O[i][j]*F_W_T[i][j];
                        }
                

                S_F_W1=S_F_W2=0;
                for(i=m+1;i<N+n+1;i++)
                        for(j=1;j<i;j++)
                        {
                                S_F_W1 += F_W_O[i][j] * F_W_T[i][j];
                                S_F_W2 += F_W_O[i][j] * F_W_O[i][j];
                                s+=F_W_T[i][j]*F_W_T[i][j];
                        }
                if ((S_F_W1>S_F_W2) || (S_F_W1==S_F_W2))
                        Lr_W=Lr_W*(a+b);
                else if (S_F_W1<(-2)*S_F_W2)
                        Lr_W=Lr_W*(a-2*b);
                     else
                        Lr_W=Lr_W*(a+b*(S_F_W1/S_F_W2));  
                
                S_F_net1=S_F_net2=0;
                for(i=m+1;i<N+n+1;i++)
                {
                        s+=F_net_T[i]*F_net_T[i];
                        S_F_net1 +=F_net_O[i] *F_net_T[i];
                        S_F_net2 +=F_net_O[i] *F_net_O[i];
                }
                if ((S_F_net1>S_F_net2) || (S_F_net1==S_F_net2))
                        Lr_ww=Lr_ww*(a+b);
                else if (S_F_net1<(-2)*S_F_net2)
                        Lr_ww=Lr_ww*(a-2*b);
                     else 
                                Lr_ww=Lr_ww*(a+b*(S_F_net1/S_F_net2));
                F_Ws1=F_Ws_O*F_Ws_T;
                F_Ws2=F_Ws_O*F_Ws_O;
                if ((F_Ws1>F_Ws2) || (F_Ws1==F_Ws2))
                        Lr_Ws=Lr_Ws*(a+b);
                else if (F_Ws1<(-2)*F_Ws2)
                        Lr_Ws=Lr_Ws*(a-2*b);
                     else   
                        Lr_Ws=Lr_Ws*(a+b*(F_Ws1/F_Ws2));
                s+=F_Ws_T*F_Ws_T;
                es=e/s;
                if ((e-Lr_W*s)<0)
                        Lr_W=Lr_W*es;
                if ((e-Lr_ww*s)<0)
                        Lr_ww=Lr_ww*es; 
                for(i=m+1;i<N+n+1;i++)
                {
                        for(j=1;j<i;j++)
                        {
                                W_O[i][j]=W[i][j];
                                W[i][j]-=Lr_W*F_W_T[i][j];
                        }
                        ww_O[i]=ww[i];
                        ww[i]-=Lr_ww*F_net_T[i];
                }
                if ((e-Lr_Ws*s)<0) 
                        Lr_Ws=Lr_Ws*es;
                Ws_O=Ws;
                Ws-=Lr_Ws*F_Ws_T;

                sum=0;
                for(i=m+1;i<N+n+1;i++)
                {
                        for(j=1;j<i;j++)
                        {
                                F_W_O[i][j]=F_W_T[i][j];
                        }
                        F_net_O[i]=F_net_T[i];
                }
                F_Ws_O=F_Ws_T;
        }
        fclose(f);
}

void synthesis(int B[30][30],int A[30][30],int n1,int n2)
{
        int k,mini,no,i,j;

        /*  Initialization */

        k=0;
        for (i=0;i<7;i++)
                for(j=0;j<7;j++)
                {
                        A[i][j]=B[i][j];
                }

/* Searching the optimal path */

        /* Calculating the Utility */
        no = n2-3-1;
        while (k!=no)
        {
                k=0;
                for(i=1;i<6;i++)
                        for(j=1;j<6;j++)
                        {
                  mini = 1 + minimum(A[i-1][j],A[i][j-1],A[i+1][j],A[i][j+1]);
                                if ((A[i][j]!=n2) && (A[i][j]!=1))
                                {
                                        if ((A[i][j]==mini) && (A[i][j]!=n1))
                                                k++;
                                        else
                                                if (mini!=n2)
                                                        A[i][j]=mini;
                                }
                        }
        }
}

/* minimum: return the minimum value */

int minimum(int s,int t,int u,int v)

{
        int mini;
        mini=0;
        mini=min(min(min(s,t),u),v);
        return mini;
}

void NET(double W[30][30],double x[30],double ww[30],
                int n, int m, int N, double Yhat[30])

{
        int i,j;
        double net;
        for (i=m+1;i<N+n+1;i++)
        {
                net=0;
                for (j=1;j<i;j++)
                {

                        net += W[i][j] * x[j];
                }
                x[i]=f(net+ww[i]);
        }
        for (i=1;i<n+1;i++)
        {
                Yhat[i]=x[i+N];
        }
}

double f(double x)
{
        double z;
        z=(1-exp(-x))/(1+exp(-x));
        return z;
}

void F_NET2(double F_Yhat,double W[30][30],double x[30],int n,int m,int 
N,double F_W[30][30],double F_net[30],double F_Ws[30],double Ws, 
 double F_x[30])
/* This subroutine calculates the F_W terms needed to adapt a fully connected */
/*      generalized MLP. It does not backpropagate through the network and it */
/*      does not permit the switching off of weights. */
{
        int i,j;
        for (i=1;i<N+1;i++)
                F_x[i]=0;
        F_x[N+n]+=F_Yhat*Ws;
        F_Ws[1]=F_Yhat*x[N+n];
        F_net[N+n]=F_x[N+n]*(1-x[N+n]*x[N+n])*0.5;
        for (j=1;j<N+n;j++)
                F_W[N+n][j]=F_net[N+n]*x[j];
        for (i=N+n-1;i>m;i--)
        {
                for (j=i+1;j<N+n+1;j++)
                {
                        F_x[i]+=W[j][i]*F_net[j];
                }
                F_net[i]=F_x[i]*(1-x[i]*x[i])*0.5;
                for (j=1;j<i;j++)
                {
                        F_W[i][j]=F_net[i]*x[j];
                }
        }
        for(i=m;i>0;i--)
                for(j=m+1;j<N+n+1;j++)
                       F_x[i]+=W[j][i]*F_net[j];

}

int min(int k, int l)
{
        int r;
        if (k>l) r=l;
        else r=k;
        return r;
}

void pweight(double Ws,double F_Ws_T,double ww[30],double F_net_T[30],
                double W[30][30], double F_W_T[30][30],int n,int N,int m)
{
        int i,j;
        for(i=m+1;i<N+n+1;i++)
                {
                        for(j=1;j<i;j++)
                        { 
                                printf("\n W[i][j] F_W_T %e %e",W[i][j],F_W_T[i][j]);
                        }
                        printf("\n ww F_net_T %e %e",ww[i],F_net_T[i]);
                }
                printf("\n Ws F_Ws_T %e %e",Ws,F_Ws_T); 

}
\end{verbatim}

\end{document}